\documentclass[10pt,journal,compsoc]{IEEEtran}
\ifCLASSOPTIONcompsoc
  \usepackage[nocompress]{cite}
\else
  \usepackage{cite}
\fi
\ifCLASSINFOpdf
  \usepackage[pdftex]{graphicx}
\else
  \usepackage[dvips]{graphicx}
\fi
\usepackage{amsmath}
\usepackage{algorithmic}
\usepackage{array}
  \usepackage[caption=false,font=footnotesize,labelfont=sf,textfont=sf]{subfig}
\usepackage{algorithm}
\DeclareMathOperator*{\argmax}{arg\,max}
\newcommand{\pluseq}{\mathrel{+}=}

\usepackage{etoolbox}
\usepackage{multirow}
\usepackage{amssymb}
\usepackage{relsize}

\usepackage{amsthm}
\theoremstyle{definition}
\newtheorem{definition}{Definition}[section]

\newcommand{\algorithmicdoinparallel}{\textbf{do in parallel}}
\makeatletter
\AtBeginEnvironment{algorithmic}{%
  \newcommand{\FORALLP}[2][default]{\ALC@it\algorithmicforall\ #2\ %
    \algorithmicdoinparallel\ALC@com{#1}\begin{ALC@for}}%
}
\usepackage{diagbox}
\makeatother
\begin{document}
\title{Scalable prediction of global online media news virality}

\author{Xiaoyan~Lu,~\IEEEmembership{Student Member,~IEEE,}
        and~Boleslaw~K.~Szymanski\footnotemark*,~\IEEEmembership{Fellow,~IEEE}% <-this % stops a space
\IEEEcompsocitemizethanks{
\IEEEcompsocthanksitem
* Corresponding author, E-mail: boleslaw.szymanski@gmail.com
\IEEEcompsocthanksitem
Xiaoyan Lu and Boleslaw K. Szymanski are with the Department of Computer Science, Rensselaer Polytechnic Institute, Troy, NY, 12180.\protect\\
%% note need leading \protect in front of \\ to get a newline within \thanks as
%% \\ is fragile and will error, could use \hfil\break instead.
}% <-this % stops an unwanted space
\thanks{Manuscript received Dec 25, 2018; revised July 11, 2018}
}

\IEEEtitleabstractindextext{%
\begin{abstract}
News reports shape the public perception of the critical social, political and economical events around the world. Yet, the way in which emergent phenomena are reported in the news makes the early prediction of such phenomena a challenging task. We propose a scalable community-based probabilistic framework to model the spreading of news about events in online media. Our approach exploits the latent community structure in the global news media and uses the affiliation of the early adopters with a variety of communities to identify the events widely reported in the news at the early stage of their spread. The time complexity of our approach is linear in the number of news reports. It is also amenable to efficient parallelization. To demonstrate these features, the inference algorithm is parallelized for message passing paradigm and tested on RPI Advanced Multiprocessing Optimized System (AMOS), one of the fastest Blue Gene/Q supercomputers in the world. Thanks to the community-level features of the early adopters, the model gains an improvement of 20\% in the early detection of the most massively reported events compared to the feature-based machine learning algorithm. Its parallelization scheme achieves orders of magnitude speedup.
\end{abstract}

\begin{IEEEkeywords}
Online Media, Information Cascades, Community Detection, Parallelization, Supercomputer
\end{IEEEkeywords}}

\maketitle

\IEEEdisplaynontitleabstractindextext

% \fi
\IEEEpeerreviewmaketitle

\IEEEraisesectionheading{\section{Introduction}\label{sec:introduction}}

\IEEEPARstart{O}{nline} media delivers news stories to the public every day. These reports shape people's perception of the ongoing social, political and economical changes around them. Although numerous events are reported in the news every hour around the globe, only a few reported events attract enormous attention of the online media - hundreds of news reports suddenly break out after the critical event happens. The burstiness of the reporting behavior in the online media makes the prediction of which events will trigger viral news challenging.

Many previous works~\cite{weng:men,zhao:erd,guil:hac} model the information diffusion as epidemics in networks - the acceptance of information is viewed as an infection of a node by infected neighbors in the network. These works assume that there exist an explicit propagation pathway which is sufficient to explain the observed information diffusion~\cite{yang:les2} - messages can only spread along the predefined edges between nodes. Although the research shows that the news reports of an event are usually confined to the geographical and cultural boundaries~\cite{lu:szy} between the global news sites, the explicit connections between pairs of individual news sites are typically unknown. Over-defining the edges by assuming that the co-reporting relationship might exist between any two news sites would inevitably result in an extremely dense network and require the number of parameters in the order of square of the number of nodes. Therefore, we focus on modeling the news sites in the online media and propose a general probabilistic framework which directly infer the affiliation of nodes with the communities. Although our model does not use the explicit network topology, the node clustering based on these affiliations matches the community structure detected by traditional community detection algorithms in the explicit network topology.
The early adopters of an information cascade, which are embedded in the so-inferred community structure, are used to predict the final cascade size.

In the global online news media, news sites usually have a preference for the content of their reports due to the local regional reach. Although many media companies' ambition is to have global market presence, most media sites have only a regional reach~\cite{hjar:1}. This regional reach phenomenon is supported by the surveys of local newscasts~\cite{gill:iyen} that demonstrates the dominance of local news. According to the geographical and cultural boundaries among the global news sites, we model clustering of news sites as community structures which are also widely observed in a variety of technological, biological and social networks~\cite{girv:new}. In the context of information cascades, the community structures play an important role in facilitating the local spread of messages~\cite{nemat:fer} because the community members are more likely to accept inputs from each other than from the outsiders. On the other hand, the community structures slow the global diffusion~\cite{kars:kiv} by trapping the news in dense regions and thus preventing global penetration. The experimental results show that, compared to machine learning models which extract point process-based features, our model exploiting the community-level signals improves the prediction accuracy by about 20\%.

We also parallelize the inference algorithm for distributed memory machines. The proposed parallelization scheme uses the standard message passing approach to exchange data between different processors. We design an asynchronous inter-core communication paradigm so that the local computations occur simultaneously with the cores exchanging data with each other. We evaluate the MPI implementation of our parallelization scheme on the petaflops class IBM Blue Gene/Q supercomputer at RPI. The algorithm gains orders of magnitude speedup inferring the parameters for the large networks, and it scales well with the number of nodes, the number of cascades, and the number of communities in a network.

In~\cite{lu:szy}, we introduced an initial work on discovery of viral cascades based on their initial period observations. This work was limited to sequential execution and the algorithm for cascade likelihood evaluation had quadratic time complexity because the probability of infection between every pair of nodes in the same cascade was computed. In contrast, our new model assigns a set of parameters to each cascade, modeling the infections of nodes by the cascades instead of the infections between every pair of nodes. Therefore, evaluating the likelihood of a cascade in our new model has a linear time complexity, and requires much less communication overhead than before when running the inference algorithm on distributed memory machines, making it scalable to large news media networks and large number of cascades.

To conclude, the major contributions of this paper are:
\begin{itemize}
    \item A new modeling approach in which information diffusion is modeled at the community level instead of the level of the individual nodes, which reduces the computational complexity of the inference algorithm.
    \item A parallelization scheme based on message passing paradigm that infers the community structure without the explicit topology of the network, which makes it scalable because it gains orders of magnitude speedup in computing the parameters for the large networks.
\end{itemize}

\section{Approach}

\subsection{Community affiliation model with cascades}

Community structures are widely observed in a variety of networks. Based on patterns of news about events spreading in the online news media, we present two basic observations below: 
\begin{itemize}
    \item Most news sites have a preference for the topics of their news coverage, e.g. politics, finance, education, military, technology or sports.
    \item The news reports are usually confined to the geographical and cultural boundaries. Although many media companies have ambitions to have global market presence, most media sites have only a regional reach~\cite{hjar:1}.
\end{itemize}

Much like many community detection methodologies \cite{blon:gui,clau:newm,lu:kuz,fort:1} which aim at discovering dense sub-graphs embedded in a network, we seek to recover such community structure for the online news sites. However, in the global and local news markets~\cite{leet:sch}, connecting every two sites reporting the same event could result in an extremely dense network. Finding community structure in such network would not only incur high computational power but would also cloud the real connections among communities. Therefore, we find the latent community affiliation of these news sites rather than drawing the edges between specific news sites. Since the underlying network topology is unknown, our probabilistic model incorporates the time points of each infection in the observed news cascades, i.e., the time of each news report. Formally, we present the definition of information cascade below.
\theoremstyle{definition}
\begin{definition}{\textbf{Information Cascade}}
A information/news cascade is a set of infections $\{(u, t_u)\}$. Each infection $(u, t_u)$ consists of a node $u$ and its infection time $t_u$.
\end{definition}

Our model assumes the news cascades happen at the community level. For a particular community of news sites, the probability of a site reporting a particular news depends on both the affiliation of this news sites with the community and the probability of the news cascade reaching its community. To formalize this idea, we define latent community below.

\theoremstyle{definition}
\begin{definition}{\textbf{Latent Community}}
A latent community is a set of nodes that, conditioned on the infection of one member, the probability that other members will get infected within a limited period is much higher than for the non-members.
\end{definition}

\begin{figure}
\includegraphics[width=7cm]{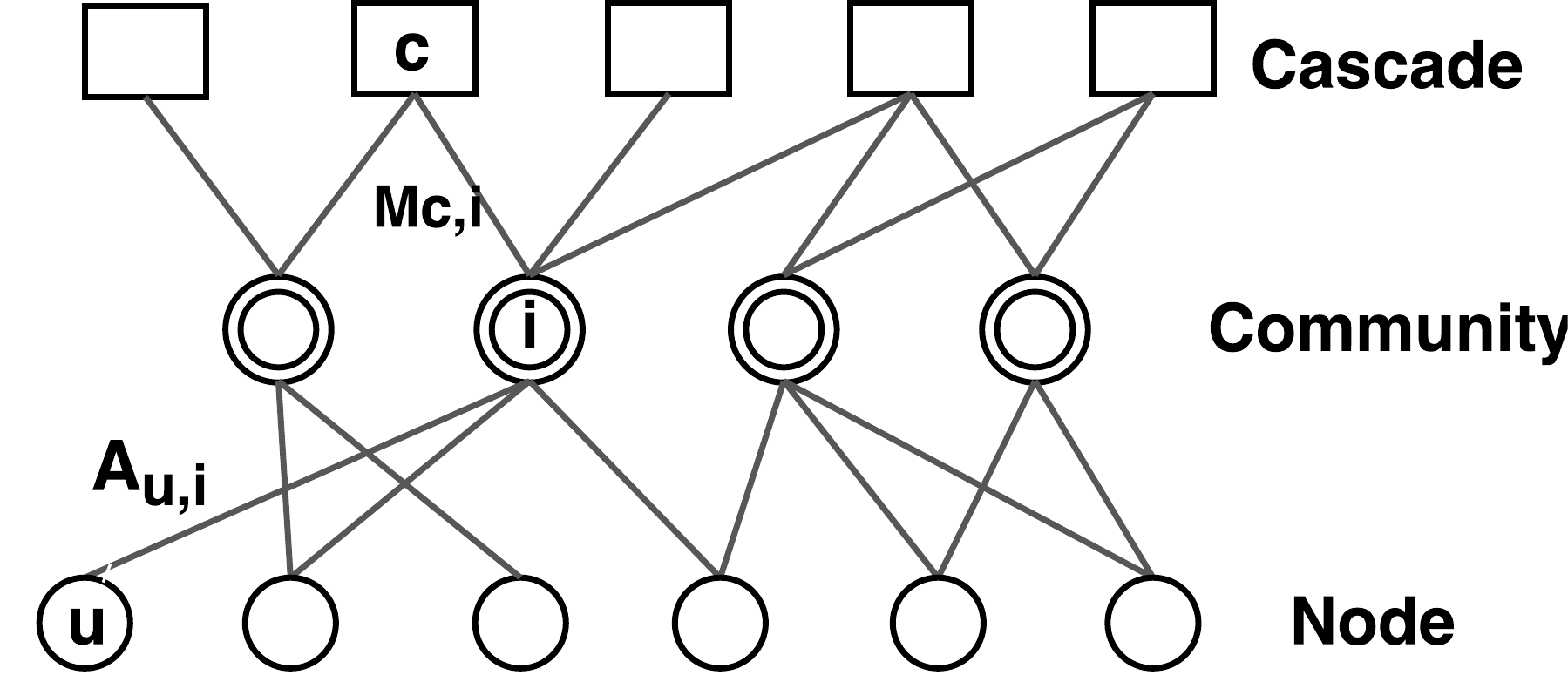}
\centering
\caption{Community affiliation model with cascades. Cascade $c$ belongs to the two leftmost communities, and node $u$ belongs to community $i$.}\label{fig:affiliation_model1}
\end{figure}

In \cite{yang:les}, authors propose a community affiliation model where each node has a probability belonging to one of the overlapping communities in the network. We extend this community affiliation model by taking into account the probability for each cascade to spread to these overlapping latent communities. As illustrated in Figure~\ref{fig:affiliation_model1}, if a cascade has a non-zero probability of spreading to a community, this cascade (square) is connected to the community (double circle); if a node (circle) belongs to a community, then this node is connected to the community as well. Our affiliation model captures two important features of cascades and communities in complex networks: (i) the community structures are highly overlapped, a node can belong to multiple communities at the same time; (ii) the members of the same community tend to accept similar information during the information cascades. 

Social theories~\cite{bara:1} suggest that the response time of human beings usually follows the exponential distribution, which explains the burstiness of social behavior in many scenarios. Our analysis of 723,037 randomly sampled news reports shows that the delays of news reports also fit the exponential distribution with a high $R^2$ score of 92\%\footnote{We consider the news reports with a delay less than 21 hours, which comprise 99\% of the sampled data from the Global Database of Events, Language, and Tone (GDELT) dataset (http://www.gdeltproject.org).}. Based on this observation, we model the response time of news sites to events by an exponential distribution.

For every single community $i$, let $A_{ui} > 0$ denote the strength of affiliation of node $u$ with it and $M_{ci} \in (0,1]$ denote the probability of the cascade spreading to community $i$. In this community, the response time of node $u$ to cascade $c$ follows the exponential distribution
\begin{equation} \label{eq:expo_ez}
t_u^c(i) \sim Exp(A_{u,i} M_{c,i})
\end{equation}
which indicates that the node $u$ gets quickly infected when the cascade $c$ is highly likely to reach a community $i$, i.e. $M_{c,i}$ is large, and the affiliation of $u$ with community $i$ is strong, i.e. $A_{u,i}$ is large.

Most real-world networks have overlapping community structures which allow a node to belong to many communities. Therefore, in our model, the response time of a node to a cascade corresponds to the minimum response time in all communities. Given a node $u$, a cascade $c$ and a total of $m$ overlapping communities, the minimum response time also follows the exponential distribution\footnote{The minimum of $m$ mutually independent random variables $X_i \sim Exp(\lambda_i)$ for $i=1,2,\dots,m$ also follows the exponential distribution, i.e. $min(X_i) \sim Exp(\sum_i\lambda_i)$.} 

\begin{equation} \label{eq:total}
\min \{ t_u^c(1),t_u^c(2),\dots, t_u^c(m) \} \sim  Exp(\sum_{i=1}^{m} A_{u,i} M_{c,i} ) 
\end{equation}
where the rate parameter is the sum of the rate parameters for each individual exponential distribution in Eq.~\ref{eq:expo_ez}.

In reality, news agencies usually have some bias for the content of information they spread~\cite{hack:1, dell:kapl, gent:shap}. Such bias can be positive - there are certain types of information a news agency favors, while it can also be negative - messages of no interest to agency's audience are ignored or intentionally blocked. It is tempting to allow the value of $M_{c,i}$ to be negative for this reason so that the affiliation with some community may delay the response time of some node, i.e. $A_{c,i} M_{c,i} < 0$. However, it could result in a negative rate parameter $\lambda$ in Eq.~\ref{eq:total} which violates the constraint $\lambda > 0$ of the exponential distribution. Hence, we smooth the rate parameter via the sigmoid function $\sigma : \mathbb{R} \to [0,1]$. In this way, given an information cascade $c$, the response time of a node $u$, $t_u^c = \min  t_u^c(i)$, draws from an exponential distribution with a rate parameter $\gamma = w \sigma( A_u \cdot M_c)$,
\begin{align} \label{eq:sigmoid}
t_u^c \sim Exp(w \sigma(A_u \cdot M_c))
\end{align}
where $\sigma(x)= \frac{1}{1+e^{-x}}$ is the sigmoid function, $w$ is a scaling parameter and $\cdot$ represents the inner product in vector space, while the $i$-th component of vector $A_u$ and $M_c$ are  $A_{ui}$ and $M_{ci}$ respectively. Using the sigmoid function here avoids any constraint for the parameters $A_u$ and $M_c$, and most importantly, it improves the robustness of the parameter estimation because the sigmoid function is differentiable at each point. 

So far, the model takes into account the participants of a cascade. However, the nodes which are not involved in a cascade also provide information about their affiliations with different communities. Since these silent nodes are equivalently important, we set a upper bound on the response time, $T \gg t_u^c $ for $\forall u$ and $\forall c$, such that all the silent nodes during the cascading propagation can be assumed to have this long response time $T$.

Given an information cascade $c=\{(u_i^c, t_i^c) | i=1,2,\dots,n_c\}$ where the $i$-th node $u_i^c$ has response time $t_i^c$, the likelihood of observing a cascade $c$ is
\begin{equation} \label{eq:casc_likelihood}
L_c = \prod_{u\in V_c} p_{u,c}(t_u^c)  \prod_{u \notin V_c}  p_{u,c}(T)
\end{equation}
where $V_c = \{u_i^c | i=1,2,\dots,n_c\}$ denotes the set of nodes involved in cascade $c$ and $p_{u,c}(t)$ is the probability density function of the exponential distribution
\begin{equation} \label{eq:pdf}
p_{u,c}(t) = w\sigma( A_{u} \cdot M_c) e^{-w\sigma( A_{u} \cdot M_c) t}
\end{equation}
Since the likelihood $L_c$ is the product of $|V|$ terms, Eq.~\ref{eq:casc_likelihood} can be too expensive to compute. But it can be approximated by using a subset of representatives $D_c$ drawn randomly from $V\setminus V_c$. This idea is similar to the negative sampling approach~\cite{mnih:teh}, which has been successfully applied to learning the distributed representation of words in documents~\cite{miko:sut,ji:sat}. The log-likelihood of an observed cascade then becomes
\begin{equation} \label{eq:neg}
\begin{array}{rcl}
   \mathcal{L}_c \approx & \sum_{u \in V_c} \log p_{u,c}(t_u^c) + \sum_{u\in D_c} \log p_{u,c}(T)
\end{array}
\end{equation}
where $D_c$ is the set of negative samples chosen for every cascade by drawing random nodes uniformly from $V$. If we fix the size of each $D_c$ as $d$, then a speedup of approximately $|V|/d$ times can be achieved.

To estimate the parameters $A_u$ for each $u$ and $M_c$ for each $c$, we maximize the likelihood which factorizes into the product of the likelihoods of $k$ cascades. This goal is equivalent to maximizing the sum of the log-likelihood of $K$ cascades
\begin{equation} \label{eq:loglikelihood}
\{\hat{A}_u\}, \{\hat{M}_c\} = \argmax_{\{A_u\}, \{M_c\}} \sum_{c=1}^{k} \mathcal{L}_c
\end{equation}
It is worth noting that the problem defined by Eq.~\ref{eq:loglikelihood} does not require explicit network topology to estimate $A_u$ and $M_c$. Instead, the input of the model is the response times of the nodes to every cascade. This is a practical setting when the underlying network topology is incomplete or hidden during the information propagation process. In addition, the parameter space $\{A_{ui},M_{ci}|\forall i,u,c\}$ in Eq.~\ref{eq:loglikelihood} does not have any restriction thanks to the adoption of the sigmoid function in Eq.~\ref{eq:sigmoid}.

\subsection{Parameter estimation}
The optimization problem in Eq.~\ref{eq:loglikelihood} is unfortunately not convex. Since the network size can be large, the optimization problem involves a large number of parameters, which makes stochastic updates more suitable than the batch methods. In addition, when some new cascade data comes in, the estimation algorithm should be able to incorporate the new cascades efficiently. For these reasons, we apply the Stochastic Gradient Ascent (SGA) method to estimate the parameters.

If we substitute Eq.~\ref{eq:neg} into Eq.~\ref{eq:loglikelihood}, the partial derivative of the objective function $F$ in Eq.~\ref{eq:loglikelihood} over a particular $M_c$ becomes 
\begin{equation} \label{eq:partial}
    \frac{\partial F}{\partial M_c} =  \sum_{u\in V_c} \frac{\partial \log p_{u,c}(t^c_u) }{\partial M_c} + \sum_{u\in D_c} \frac{\partial \log p_{u,c}(T) }{\partial M_c}
\end{equation}
which is a weighted sum of the terms in the form of $\frac{\partial \log p_{u,c}(t) }{\partial M_c}$. Given the value of $t$, $p_{u,c}(t)$ depends only on $A_u$ and $M_c$. The partial derivative of $\log p_{u,c}(t)$ over $M_c$ can be computed using $A_u$ and $M_c$
\begin{equation} 
    \frac{\partial \log p_{u,c}(t)}{\partial M_c} = \big[1 - \sigma( A_u \cdot M_c) w t\big] \big[1-\sigma( A_u \cdot M_c)\big] A_u
\end{equation}
Here we need the $A_u$ for $\forall u \in V_c \cup D_c$ to update $M_c$ according to the gradient in Eq.~\ref{eq:partial}. Similarly, the partial derivative of the objective function $F$ in Eq.~\ref{eq:loglikelihood} over $A_u$ can be computed using the $M_c$ for cascades $c$ such that $u \in V_c \cup D_c$
\begin{equation}
\frac{\partial F}{\partial A_u} =  \mathlarger{\sum_{c:u\in V_c}} \frac{\partial \log p_{u,c}(t^c_u) }{\partial A_u} + \mathlarger{\sum_{c:u\in D_c}} \frac{\partial \log p_{u,c}(T) }{\partial A_u}
\end{equation}
where the term $\frac{\partial \log p_{u,c}(t)}{\partial A_u}$ depends on $A_u$ and $M_c$ only
\begin{equation} 
    \frac{\partial \log p_{u,c}(t)}{\partial A_u} = \big[1 - \sigma( A_u\cdot M_c) w t\big] \big[1-\sigma( A_u\cdot M_c)\big] M_c
\end{equation}
In this way, the parameters can be updated in a pair-wise manner: we fix $A_u$ for all $u$ and update all the $M_c$s, and then fix all $M_c$s to update $A_u$s in every SGA iteration. The SGA updates can operate on a bipartite graph where node $u$ and cascade $c$ are connected if $u\in V_c \cup D_c$ (cf. the example shown in leftmost diagram in Figure~\ref{fig:pll}). To update $A_{u}$, each cascade $c$ propagates the corresponding parameters $M_c$ through the links in the bipartite graph to the node $u$. Then node $u$ calculates the partial derivative $\frac{\partial \log p_{u,c}(t) }{\partial A_{u}}$ for every connected cascade $c$ and updates $A_{u}$ accordingly. Similarly, the nodes can propagate the parameters $A_u$s through the links in this bipartite graph towards each relevant cascade $c$, and $M_{c}$ can be updated using the $A_u$ it receives. 

The pseudo code is shown in Algorithm~\ref{algo:1}.
The \textbf{time complexity} of each SGA iteration here is linear in the number of edges in the bipartite graph. This is because the loop in lines 8-19 iterates over all the nodes $u$ connected with each cascade $c$, which corresponds to visiting every edge of the bipartite graph exactly once. Similarly, the lines 20-31 also visit every edge in the bipartite graph once. In the news dataset, the number of edges of the bipartite graph is defined by the number of news reports, because every report connects a news site to a news cascade, plus the pair of negative samples edges $(u, c)$ for $u\in D_c$ with $|D_c|=d$ fixed as a constant. Hence, the time complexity of each SGA iteration is linear in the number of edges in the bipartite graph. The time complexity of Algorithm~\ref{algo:1} is also determined by the number of SGA iterations. In practice, this only involves tens of iterations before the derived $A_u$ and $M_c$ vectors become stable. Thus, this number can be treated as a constant. Therefore, the Algorithm~\ref{algo:1} has a linear time complexity in the number of bipartite graph edges, i.e. the number of news reports.

\begin{algorithm}
\caption{SGA Algorithm using a Single Processor}\label{algo:1}
\begin{algorithmic}[1]
\STATE $\alpha = \text{the SGA stepsize}$
\FOR {each cascade $c$} 
\FOR {each node $u \in V_c$}
\STATE $t_u^c = \text{infection delay of node $u$ in cascade $c$}$
\ENDFOR
\ENDFOR
\FOR {each SGA iteration}
\FOR {each cascade $c$} 
\FOR {each node $u \in V_c \cup D_c$}
\IF {$u \in V_c$}
\STATE $t= t_u^c$
\ELSIF{$u \in D_c$}
\STATE $t= T$
\ENDIF
\FOR {$i = 1,2,\dots,m$}
\STATE $A_{ui} \pluseq \alpha \frac{\partial \log p_{u,c}(t) }{\partial A_{ui}}$
\ENDFOR
\ENDFOR
\ENDFOR
\FOR {each node $u \in V$}
\FOR {each cascade $c$ such that $u \in V_c \cup D_c$}
\IF {$u \in V_c$}
\STATE $t= t_u^c$
\ELSIF{$u \in D_c$}
\STATE $t= T$
\ENDIF
\FOR {$i = 1,2,\dots,m$}
\STATE $M_{ci} \pluseq \alpha \frac{\partial \log p_{u,c}(t) }{\partial M_{ci}}$
\ENDFOR
\ENDFOR
\ENDFOR
\ENDFOR
\end{algorithmic}
\end{algorithm}

\subsection{Parallelization for distributed memory machines}
In practice, the input network size can be large so to speed up computation we offer parallelization of the parameter estimation for our model. Since the SGA algorithm is inherently sequential, many works including~\cite{ji:sat,miko:sut} use the Hogwild! framework~\cite{rech:re} attempting to parallelize the SGA algorithm on shared memory machines. The Hogwild! framework ignores the write-write conflicts caused by parallel updates on the same parameter as long as one unique processor can complete its writing operation. The quality of the results produced by SGA algorithm is guaranteed by the property that such conflicts are sparse enough. In the information diffusion context, however, such data sparsity is uncertain. To handle the contentions between processors, we propose a scalable parallelization scheme based on message passing paradigm~\cite{grop:lus}.

\begin{figure}[t]
\includegraphics[width=9cm]{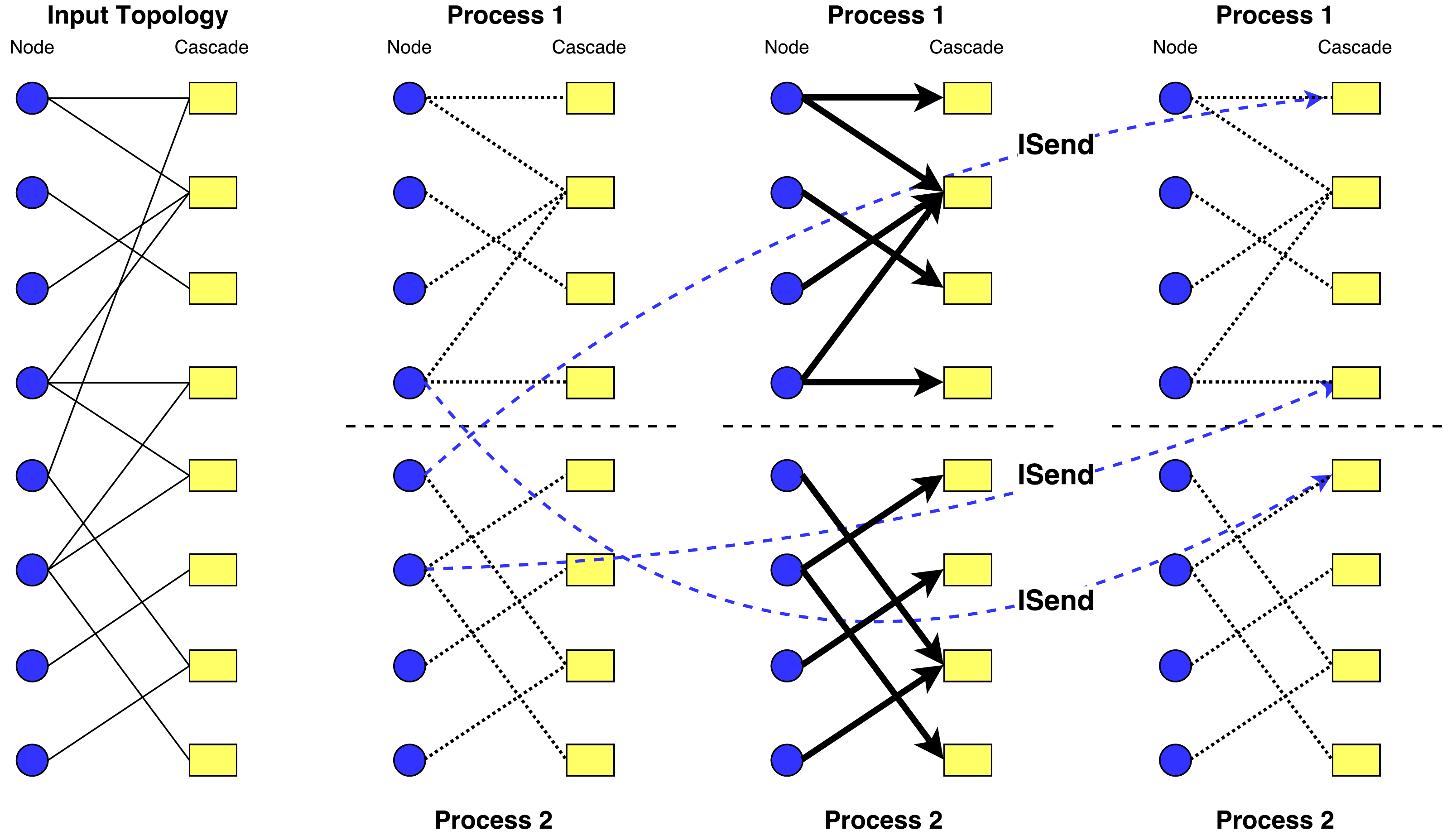}
\centering
\caption{Illustration of the parallelization scheme using two processors. The parameters propagate back and forth between nodes and cascades iteratively. If the connected node and cascade are in the different processors, the parameters are sent via asynchronous communication, i.e. the blue dashed lines marked by ``ISend''. At the same time, the local parameter propagation occurs within each processor, i.e. the bold black arrows in the middle. This figure illustrates the parameter propagation from node layer to cascade layer. The parameter propagation from cascade layer to node layer is conducted in a similar manner.}\label{fig:pll}
\end{figure}

The parallelization scheme is illustrated in Figure~\ref{fig:pll}. Consider eight nodes (blue) involved in eight cascades (yellow) in this toy example. The bipartite graph connects every pair of associated nodes and cascades in the SGA updates. These nodes and cascades are then distributed in this example to two processors. The first processor owns the upper four nodes and upper four cascades while the remaining nodes and cascades are assigned to the second processor. Each processor creates private memory space for the parameters of the nodes and cascades it owns. Much like Algorithm~\ref{algo:1}, the SGA algorithm propagates parameters back and forth between nodes and cascades, the only difference here is that the propagation between the cascades and nodes owned by different processors requires inter-core communication.

During every SGA iteration, each node $u$ propagates its $A_u$ to all the connected cascades in the bipartite graph. If these cascades are located in the same processor which owns node $u$, then this propagation operation is done locally. Otherwise, $A_u$ is sent \textbf{asynchronously} to the processor owning node $u$. The pseudo code for the parallelization scheme is shown in Algorithm~\ref{algo:2} and Algorithm~\ref{algo:3}. Without loss of generality, we consider updating $M_c$s using $A_u$s, i.e. the line 14 in Algorithm~\ref{algo:2}. One SGA iteration consists of the following three phases:
\begin{itemize}
\item (a) Message passing: Every processor sends the $A_u$s it owns to the target processors which require those $A_u$s to update their corresponding $M_c$s, cf. the lines 1-3 in Algorithm~\ref{algo:3}.
\item (b) Local updates: Every processor updates the $M_c$ it owns using the gradients $\frac{\partial \log p_{u,c}(t) }{\partial M_{c}}$ if it also owns $A_u$, cf. the lines 4-6 in Algorithm~\ref{algo:3}. 
\item (c) Remote updates: After all the remote $A_u$s sent in phase (a) have been received, each processor updates $M_c$ using the gradients $\frac{\partial \log p_{u,c}(t) }{\partial M_{c}}$ whose computation requires some of the received $A_u$s, cf. the lines 8-10 in Algorithm~\ref{algo:3}.
\end{itemize}
Note that each processor should conduct the local updates prior to the updates which require remote data from other processors, i.e. the phase (b) occurs before the phase (c). In this way, the local computation time and inter-core communication time overlap with each other, improving the parallelization efficiency. 

Figure 2 illustrates this parallelization scheme. The three phases described above are represented by the three diagrams following the first diagram showing the initial stage in Figure~\ref{fig:pll}. The parameters propagate back and forth between node layer and cascade layer iteratively. If the connected node and cascade are in the different processors, the parameters are sent via asynchronous communication, i.e. the blue dashed lines labeled ``ISend''. At the same time, the local parameter propagation occurs within each processor, as indicated by the bold black arrows in the middle. Using the same protocol, we can update $A_u$s using the values of $M_c$s.

After distributing nodes and cascades to the processors, each processor creates its private memory space for the $A_u$s and $M_c$s it owns and ghost memory space for those $A_u$s and $M_c$s connected to the nodes or cascades in the bipartite graph, but owned by other processors. In every SGA iteration, once the ghost memory is filled with the received data, it will not be written again. For example, a processor owns two nodes $u$ and $v$, both involved in a cascade $c$ which is owned by another remote processor. To update $A_u$ and $A_v$, $M_c$ is sent asynchronously to this local processor. But $M_c$ will be sent only once regardless of the number of associated nodes owned by the local processor, because $M_c$ can be shared by the updates of $A_u$ and $A_v$. This optimization is similar to the combiner applied to the message queues in Pregal-like parallel graph processing systems~\cite{male:aus,aver:1} to avoid sending duplicated messages to the same target processor.

\begin{algorithm}
\caption{Parallelized SGA Algorithm (Distributed Memory Machines)}\label{algo:2}
\begin{algorithmic}[1]
\FOR {each cascade $c$}
\STATE $proc(c) = $ ID of the processor storing $M_c$
\ENDFOR
\FOR {each node $u$}
\STATE $proc(u) = $ ID of the processor storing $A_u$
\ENDFOR
\FORALLP {each processor $p$}
\STATE $U_p = $ IDs of the nodes owned by processor $p$
\STATE $C_p = $ IDs of the cascades owned by processor $p$
\ENDFOR
\FOR {each SGA iteration}
\FORALLP {each processor $p$}
\STATE Call Algorithm~\ref{algo:3} to update $M_c$s using $A_u$s
%\LineComment{Update $A_u$s using $M_c$s}
%\LineComment{Update $A_u$s using $M_c$s}
\ENDFOR
\FORALLP {each processor $p$}
\STATE Call Algorithm~\ref{algo:3} to update $A_u$s using $M_c$s similarly
%\LineComment{Update $A_u$s using $M_c$s}
%\LineComment{Update $A_u$s using $M_c$s}
\ENDFOR
\ENDFOR
\end{algorithmic}
\end{algorithm}

\begin{algorithm}
\caption{Parallelized SGA Updates of $M_c$s using $A_u$s (Distributed Memory Machines)}\label{algo:3}
\begin{algorithmic}[1]
%\STATE \COMMENT{\it 1.a Message passing}
\FOR {each $(u,c)$ s.t. $u \in U_{p}$, $c \notin C_p$ and $u \in V_c \cup D_c$}
\STATE Send $A_u$ to $proc(c)$ asynchronously
\ENDFOR
%\STATE \COMMENT{\it 1.b Local updates}
\FOR {each $(u,c)$ s.t. $u \in U_{p}$, $c \in C_p$ and $u \in V_c \cup D_c$}
\STATE Call Algorithm~\ref{algo:1} to update $M_c$ using $A_u$.
\ENDFOR
\STATE Wait for asynchronous receive requests until done
%\STATE \COMMENT{\it 1.c Remote updates}
\FOR {each $(u,c)$ s.t. $u \notin U_{p}$, $c \in C_p$ and $u \in V_c \cup D_c$}
\STATE Call Algorithm~\ref{algo:1} to update $M_c$ using $A_u$.
\ENDFOR
\end{algorithmic}
\end{algorithm}
\subsection{Forecast viral cascade via its early adopters} \label{sec:forecast}
Our aim is to forecast the viral information cascade. From the historical cascades, the proposed model estimates the $A_u$ vector for each $u$ according to the observed response times. Using these $A_u$ vectors of the initially infected nodes, we seek to predict the behavior of future cascades.

Suppose a set of so-called early adopters have been infected within a limited time period. One basic observation is that, once the contagion reaches a community member, the probability that other members get infected increases. Therefore, we can make use of the infected node's local neighborhood to predict the infection future. Since our model presents every node $u$ by a vector $A_u$ in the latent space, it is easy to find the neighbors which are close to node $u$ by measuring the Euclidean distances between them. As the contagion is likely to spread fast in the dense areas, the number of neighbors within a certain range can be used as an indicator of future infections. Hence, we count the number of neighbors that are located within a certain range from the infected node $v$, and arrange these values in a vector $K$ whose $i$-th component is defined as
\begin{equation} \label{eq:prediction_}
K_i = |\{u|\|A_u - A_v\|_2 < r_i\}|
\end{equation}
where $r_i$ is the radius of the $i$-th neighborhood of node $v$ and $\|\cdot \|_2$ denotes the Euclidean norm.

\begin{figure}
    \centering
    \includegraphics[width=5cm]{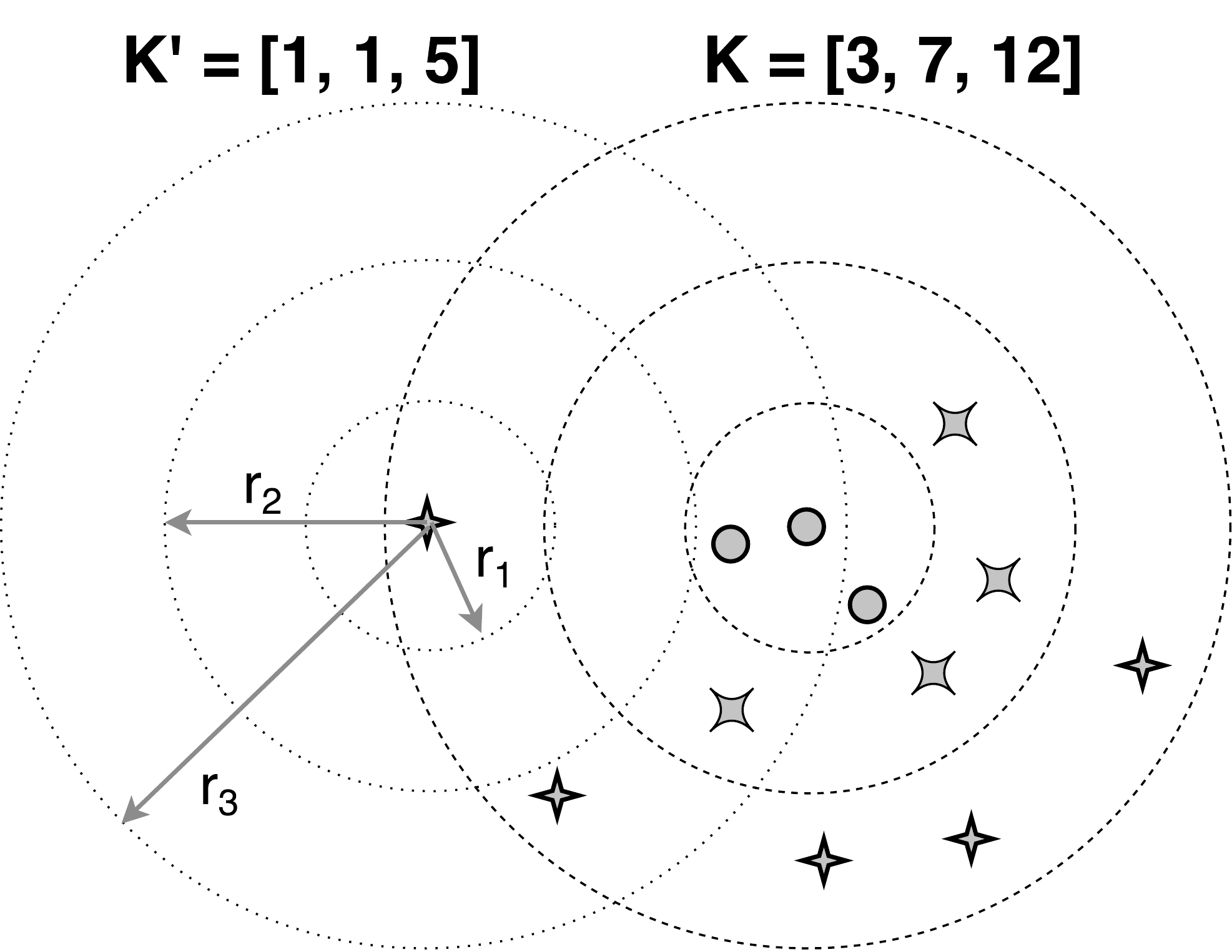}
    \caption{Illustration of the local neighborhoods in the vector space. The concentric circles mark the neighborhoods with different radii from the center. The proposed method counts the number of nodes located in each circle and arrange these values in a vector. Vector $K$ shows the numbers of nodes for $r_1$, $r_2$ and $r_3$ radii drawn from the figure center while $K'$ is shown for the circles centered at the asterisk on the left.}
    \label{fig:hood}
\end{figure}

Figure~\ref{fig:hood} demonstrates how to compute the $K$ vector of an infected node. Suppose the infected nodes are located at the centers of the different circles. Their neighborhoods are marked by the circles which have the radii $r_1$, $r_2$ and $r_3$ respectively. In Figure~\ref{fig:hood}, the node at the right center has 3 neighbors inside the small circle, 7 neighbors inside the medium circle and 12 neighbors inside large circle, resulting in the vector $K=[3,7,12]$. The leftmost node has only 1 neighbor, i.e. itself, inside both the small and medium circle and 5 neighbors inside the large circle, resulting in the vector $K'=[1,1,5]$. Intuitively, we can tell from $K$ and $K'$ that the right three neighborhoods allow faster growth of infections than the left ones. 

Given a set of early adopters which get infected within a limited time period, we can count the total number of \textbf{unique} neighbors in their local neighborhoods within different radii similarly. Note that if a node is in the $i$-th neighborhood of two early adopters, this node is only counted once in the $i$-th component of $K$. Finally, The numbers of neighbors, presented as the components of a multiple-dimensional vector $K$, are fed to a machine learning model to predict the final size of a cascade. The specific experimental configuration is presented in more detail in Section~\ref{set:virality}.

\section{Related Work}
In this section, we review the related works and summarize the relationship between our model and these works.

\subsection{Network Embedding}
Network embedding aims at representing the graph structures by the low-dimensional vectors so that they can be exploited by machine learning models. Compared to the classical dimension reduction algorithms such as multidimensional scaling (MDS)~\cite{cox:cox}, IsoMap~\cite{tene:de}, Laplacian eigenmap~\cite{belk:niy} for which time complexity is in the order of square of the number of nodes, many recent works~\cite{pero:alr,grov:les,tang:qu} take advantage of the softmax-based objective function to efficiently learn the representation of network nodes in a distributed manner~\cite{miko:sut}. Our work uses the similar approach to avoid the quadratic time complexity of the inference algorithm.

\subsection{Community Detection}
Community structures are widely observed in a variety of technological, biological and social networks. In the context of information diffusion, ~\cite{weng:men} shows that communities structures are important for the prediction of viral cascades in social networks, however, it relies on traditional community detection algorithms which require the explicit network topology. Recent works~\cite{psor:rob,wang:li,yang:les} adopt the non-negative matrix factorization approach to detect communities. In these works, the cells in factorized matrices indicate the affiliation of nodes with the communities, and the inner product of two nodes' vector representations corresponds to the probability of observing an edge between them. Based on this framework, we extend the community affiliation model to consider information cascades and negative affiliations so that the community-preserving vector representation can be efficiently obtained without the explicit network topology.

\subsection{Parallel Graph Processing}
Inspired by Valiant's Bulk Synchronous Parallel (BSP) model~\cite{vali:1}, Pregal~\cite{male:aus} conducts a sequence of iterations, called supersteps, to support efficient large graph processing. In Pregal, the computation involving individual nodes in a network would occur in parallel during every superstep, while the communication across different network nodes occurs at the same time. The safety of the parallel algorithm is ensured because a new superstep starts after the communication of the previous superstep is done. GraphLab~\cite{low:bickson} also introduces a similar approach in which the synchronization between network nodes happens between ``superstep''s. Our proposed parallelization design adopts this approach to speedup the inference algorithm, while avoiding the contention between processors.

\section{Experimental results}
In this section, we evaluate the proposed model and the parallelized inference algorithm using both synthetically generated cascades as well as the real global news reports data. Due to the lack of propagation topology in the global news media, we compare the communities discovered by our model with the communities detected by several state-of-the-art algorithms and the predefined communities in the synthetic networks. In our experiments of virality prediction, we focus on the classification of the events most reported in the news and present its accuracy measured by F1 score.

\subsection{Datasets}
\subsubsection{Synthetic cascades} \label{sec:syn_cas}
We simulate cascades in the synthetic networks generated by the Stochastic Block Model (SBM)~\cite{moss:nee} where the community structures are pre-defined. We choose Stochastic Block Model (SBM) as the random graph model to generate network topology because it provides well-defined community structures and is computationally efficient therefore suitable for generating large networks~\cite{bata:bran}. Given the network $G=(V,E)$, the simulation of cascading process is based on the Independent Cascade (IC) model~\cite{kemp:kle}. In the IC model, the infection time of a node is the earliest time when the first neighbor infects it, i.e. a node can only be infected once. After a node gets infected, it starts to spread the contagion to its uninfected neighbors. Compared to the linear threshold model in which the diffusion process unfolds deterministically, the IC model considers information diffusion a probabilistic process: if nodes $u$ and $v$ are connected and node $u$ is infected, then 
in every discrete step, node $u$ infects node $v$ with probability $p_{u,v}$. As an extension of the IC model, the infection delay can be modeled as the continuous time~\cite{gome:les}. Given the infection time of the $r$ neighbors of node $v$, the infection time of $v$ can be expressed as
\begin{align}
   &t_v = \min{ t_{u_1}, t_{u_1},\dots,t_{u_r}} \\
   &t_{u_i} \sim \mathcal{K}(\alpha_{u_i,v}) \quad \text{for  } i = 1,2,\dots,r
\end{align}
where $\alpha_{u_i,v}$ is a parameter associated with the edge $(u,v)$ in the propagation network and $\mathcal{K}()$ is the distribution of infection delays. In our experiments, $\mathcal{K}()$ is set as the exponential distribution which is observed in many social dynamics~\cite{bara:1}, and we set $\forall (u,v)\in E$: $\alpha_{u,v}=1$ for simplicity. In theory, the IC model supposes the entire network will be infected given a sufficiently long period. Since news cascades have very limited time span, in our experiments the simulation of every cascade happens within a predefined observation window~\cite{gome:les}.

\subsubsection{Data about global news of events}
The Global Database of Events, Language, and Tone (GDELT)\footnote{http://www.gdeltproject.org/}~\cite{leet:sch} project records the news reports of thousands of news sites around the world. It provides the translation of 65 languages into English and identifies the same events reported by different news sites. The dataset is currently available on Google Cloud platform. 

Since the GDELT dataset has a bias towards US domestic news, we choose the most active 2000 news sites for each country and 500 random events reported in news between July 1st, 2017 and July 19th, 2017 in the corresponding region. The sampled dataset consists of 19795 news sites and 26752 events reported in the news, where every event is reported by 27 news sites on average. Although the GDELT dataset does not indicate the connections between any pair of news sites, \cite{lu:szy} found that reports of an event are usually confined to the geographical and cultural boundaries of the event, which matches our model's assumption that information cascades are likely to happen inside communities.

\subsection{Alignment of community structures with node vectors' clustering}
Our model produces the $A_u$ vector for each node $u$ in the network. If these $\{A_u\}$ vectors preserve the community structures of the news media network well, their clustering should match the community structure embedded in the explicit network topology, because the members of a community have similar $A_u$s.

Given a synthetic SBM network, we simulate the cascades as described in Section~\ref{sec:syn_cas}. Our model then infers the $\{A_u\}$ vectors by these cascades. We compare the node clustering\footnote{For clarity, the set of nodes detected in a network is called a community and the set of nodes clustered by their vector representations is called a cluster.} of these $\{A_u\}$ vectors with the ground truth partition of the SBM network and the community structures discovered by traditional community detection algorithms from the explicit topology. The alignment between them indicates the $\{A_u\}$ vectors produced by our model preserve the community structures.

More specifically, the K-means clustering algorithm~\cite{macq:1} is executed on the so-inferred $\{A_u\}$ vectors to derive the node clustering. The similarity between the node clustering and the contrastive partitioning of the network are measured by the Adjusted Mutual Information (AMI) and Adjusted Rand Score (ARS) which are widely used to evaluate community detection performance.

\textbf{Adjusted Mutual Information} (AMI)~\cite{vinh:epp} which is defined as
\begin{equation}
AMI(U,Q)= \frac{MI(U,Q) - E[MI(U,Q)]}{max\{H(Q), H(U)\} - E[MI(U,Q)]}
\end{equation}
where $U=\{u_i\}$ is the node clustering, each set $u_i$ contains the nodes in a single cluster and $Q=\{q_i\}$ is the contrastive partition of the SBM network; the entropy associated with the partition $Q$ is defined as
\begin{equation}
H(Q) = -\sum_{q_i \in Q} p(q_i) \log{p(q_i)}\\
\end{equation}
and the mutual information (MI) between $U$ and $Q$ is defined as
\begin{equation}
MI(U, Q) = \sum_{u_i \in U}\sum_{q_i \in Q} p(u_i, q_j) \log{\frac{p(u_i,q_j)}{p(u_i)p(q_j)}}\\
\end{equation}
where
\begin{equation}
    p(q_i) = \frac{|q_i|}{|V|} \quad \quad
    p(u_i,q_j) = \frac{|u_i \cap q_j|}{|V|}
\end{equation}

\textbf{Adjusted Rand Score} (ARS)~\cite{hube:ara} which computes the similarity by comparing all pairs of nodes that are assigned to the same or different communities in partitions $U$ and $Q$
\begin{align}
\begin{split}
& ARS(U,Q) = \\
& \frac{  \sum_{ij}{\binom{|u_i \cap q_j|}{2}} - \frac{ [\sum_i {\binom{|u_i|}{2}} \sum_j{ \binom{|q_j|}{2} }   ] } { \binom{|V|}{2} } } {  \frac{1}{2} [\sum_i \binom{|u_i|}{2} + \sum_j \binom{|q_j|}{2} ]  - \frac{ [\sum_i {\binom{|u_i|}{2}} \sum_j{ \binom{|q_j|}{2} }   ] } { \binom{|V|}{2} } }
\end{split}
\end{align}

\begin{figure}
    \centering
    \includegraphics[width=8cm]{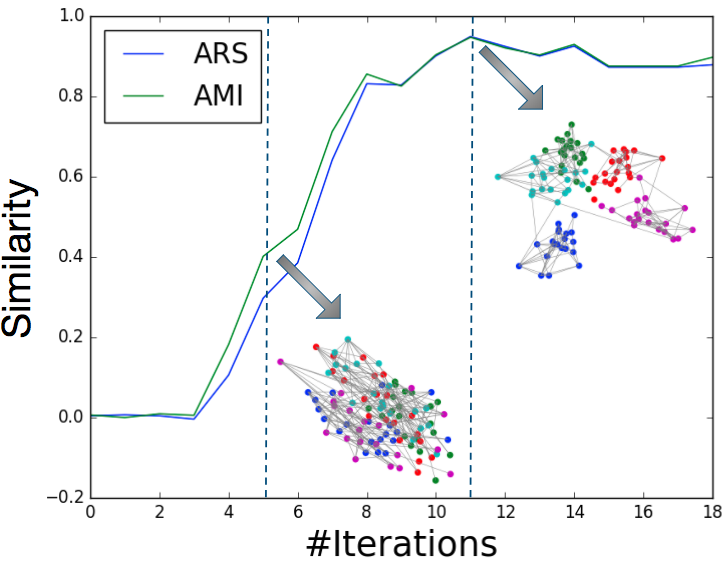}
    \caption{The similarity between the node clustering of the $\{A_u\}$ vectors at each SGA iteration of Algorithm~\ref{algo:1} and the ground truth partition of the SBM network.}
    \label{fig:alignment_figure}
\end{figure}

\begin{table}[]
\caption{The pairwise similarities between the communities detected by the state-of-the-art community detection algorithms, the node clustering of the $\{A_u\}$ vectors produced by our model and the ground truth partition of the SBM network. The entries below and above the main diagonal represent the Adjusted Rand Score (ARS) and Adjusted Mutual Information (AMI) respectively. FG: Fast Greedy algorithm~\cite{clau:newm}, LE: leading eigenvector method~\cite{new:1}, LP: label propagation algorithm~\cite{ragh:alb}, ML: multilevel algorithm~\cite{blon:gui}. Our model produces node embeddings whose clustering aligns well with the ground truth communities, even outperforming some community detection baseline methods.}\label{tab:alignment_table}
    \centering
    \begin{tabular}{|c|cccccc|}
   \hline
   \diagbox{ARS}{AMI} & FG & LE & LP & ML &  \begin{tabular}{@{}c@{}}Our \\ Model\end{tabular} &  \begin{tabular}{@{}c@{}}Ground \\ Truth\end{tabular}\\
   \hline
FG &  & 0.858 & 0.833 & 0.943 & 0.881 & 0.933 \\
LE & 0.873 &  & 0.807 & 0.867 & 0.795 & 0.837 \\
LP & 0.864 & 0.829 &  & 0.835 & 0.773 & 0.804 \\
ML & 0.929 & 0.881 & 0.868 &  & 0.922 & 0.949 \\
Our Model & 0.869 & 0.820 & 0.817 & 0.936 &  & 0.930 \\
Ground Truth & 0.939 & 0.865 & 0.852 & 0.963 & 0.925 & \\
\hline
    \end{tabular}
\end{table}

Figure~\ref{fig:alignment_figure} shows the growth of the similarity between the node clustering of the $\{A_u\}$ vectors at each SGA iteration and the ground truth partition of the SBM network. The SBM network has 100 nodes and 5 communities of size 20, 190 edges connects nodes in the same community and 12 edges are across different communities. A total of 100 cascades are simulated, each involves 12.5 infections on average. The dimension of $A_u$ is $m=10$. After each SGA iteration of Algorithm~\ref{algo:1}, we compute a $100\times 100$ distance matrix with the $(u,v)$ entries being the Euclidean distances between the latest updated vectors $A_u$ and $A_v$. The plots for two networks in Figure~\ref{fig:alignment_figure} are made by the MDS algorithm~\cite{cox:cox} which places each node in two-dimensional space such that the derived between-node distances are preserved as well as possible. In other words, the MDS coordinates preserve the nodes' pairwise distances in the high-dimensional space of $\{A_u\}$. At the 5th iteration, the ARS and AMI scores are around 0.4, the nodes' MDS coordinates do not reflect their ground truth communities represented by the color. When the 11th SGA iteration is done, the ARS and AMI scores become greater than 0.9, and the nodes' MDS coordinates match the community structures very well. Figure~\ref{fig:alignment_figure} shows that, as the inference algorithm proceeds, the $\{A_u\}$ vectors start to preserve the community structures, even though our model takes only the infection delays in the cascades as input, but not the the explicit network topology.

In addition, we compare the node clustering of $\{A_u\}$ vectors at the 15th iteration with the ground truth partition of the SBM network and community structures detected by the state-of-the-art algorithms such as Fast Greedy algorithm~\cite{clau:newm}, leading eigenvector method~\cite{new:1}, label propagation algorithm~\cite{ragh:alb} and multilevel algorithm~\cite{blon:gui}. Table~\ref{tab:alignment_table} shows that the alignments between them are good, and that the node clustering of $\{A_u\}$ vectors are more similar to the ground truth partition than the community structures detected by some state-of-the-art community detection algorithms. In Table~\ref{tab:alignment_table}, each entry indicates the similarity of the communities produced by a particular pair of methods. All the entries below the main diagonal correspond to the ARS scores and entries above correspond to the AMI scores. As the ARS and AMI scores indicate, our model produces meaningful node embeddings because the clustering of these node vectors aligns well with the ground truth communities. The community structure obtained by clustering node vectors is even closer to the ground truth than are the community structures detected by the baseline methods that include leading eigenvector method (LE) and label propagation algorithm (LP) are. Our model produces node embeddings whose clustering aligns well with the ground truth communities, outperforming some community detection baselines. In addition, our model does not use the topology of the SBM network like the baseline algorithms do, instead it only accesses the cascades data, which explains why Fast Greedy algorithm (FG) and multilevel algorithm (ML) performs better than our model in terms of community detection. Finally, it should be noted that we choose the number of clusters as 5 for the K-means algorithm here. However this number should be actually systematically selected. We leave the selection of the proper number of clusters for future work.

We also test our algorithm for large SBM networks with the dimension of resulting $A_u$ being 200. In these experiments, there are 100 predefined communities in the SBM network, each containing 100 nodes. Every node is connected to 8.8 nodes in the same community and 1.2 nodes in the other communities on average. And we simulate 10K cascades using the continuous time IC model. As shown in Table~\ref{tab:1}, as the number of processors increases, the AMI and ARS metrics are consistently above 0.98 and 0.94 respectively, which indicates the resulting $\{A_u\}$ preserves the community structure in the network. The distance matrices of the first 500 nodes are also shown in Table~\ref{tab:1}. In the distance matrix, the distance between two nodes $u$ and $v$ is defined as the Euclidean distance between vectors $A_u$ and $A_v$ and this value is visualized by the color brightness in the heatmap, brighter the color longer the distance. Each dense module in this matrix comprises 100 nodes and matches the predefined SBM community very well. As illustrated by the visualized pair-wise nodes distance matrices, the number of processors does not change the high quality of $\{A_u\}$ as the resulting vectors preserve the community structure of SBM networks in all cases. Our model does not use the topology of the SBM network, instead it only accesses the response times of the nodes to different cascades, yet the community structure can still be accurately recovered from these response times.

%SBM_10K nodes, 10K cascades, D=200 for Table 1
\begin{table}[ht]
  \centering
  \caption{Quality metric of the detected communities on synthetic SBM networks with 10K nodes. ARS: Adjusted Rand Score; AMI: Adjusted Mutual Information.}
  \begin{tabular}{ccccc}
    \hline
    \#Processors & {\bf 1} & {\bf 4} & {\bf 16} & {\bf 64}\\
    \hline
    ARS & 0.9588 & 0.9480 &	0.9444 & 0.9704\\
    AMI & 0.9858 & 0.9814 & 0.9808 & 0.9888\\
    \hline
  \end{tabular}
  \label{tab:1}
\end{table}

\begin{figure}[t]
\centering
\begin{tabular}{cc}
  \subfloat[1 processor]{\label{fig:c1}%
  \includegraphics[width=4.3cm]{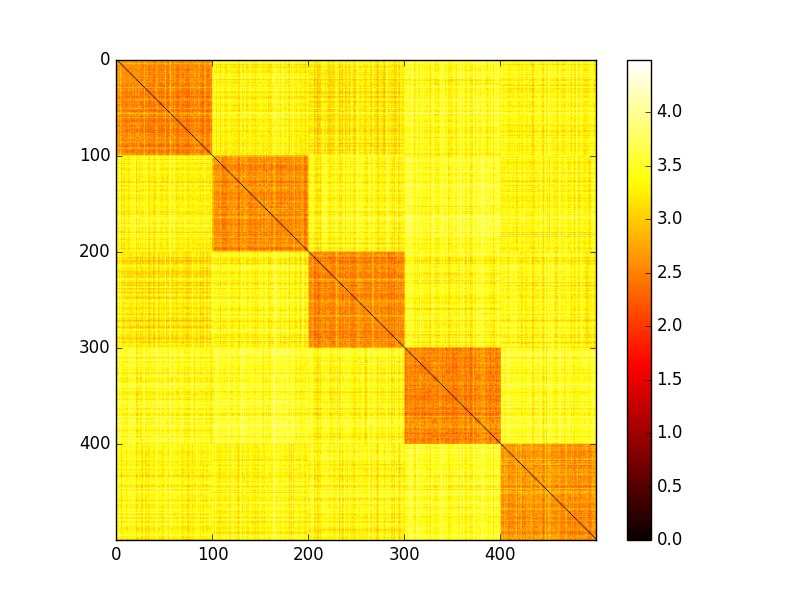}
  } &
  \subfloat[4 processors]{\label{fig:c2}%
  \includegraphics[width=4.3cm]{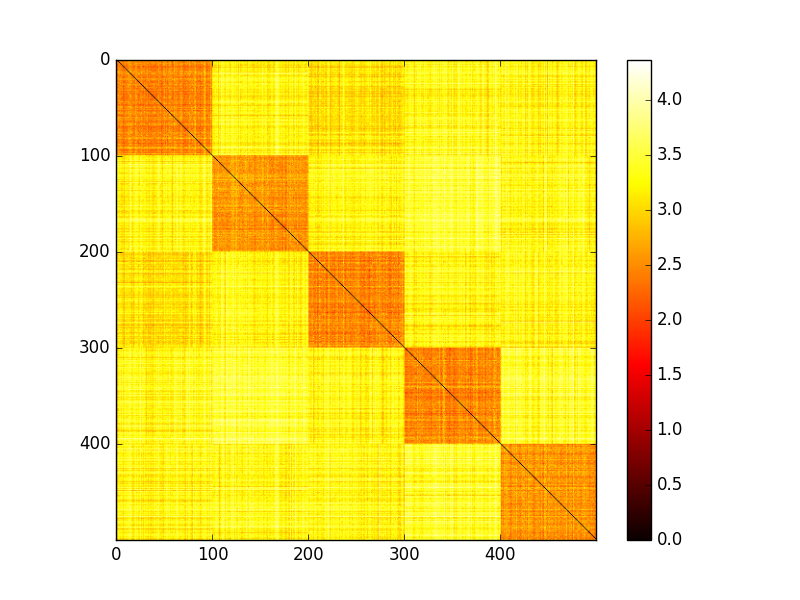}
  } \\
  \subfloat[16 processors]{\label{fig:c3}%
  \includegraphics[width=4.3cm]{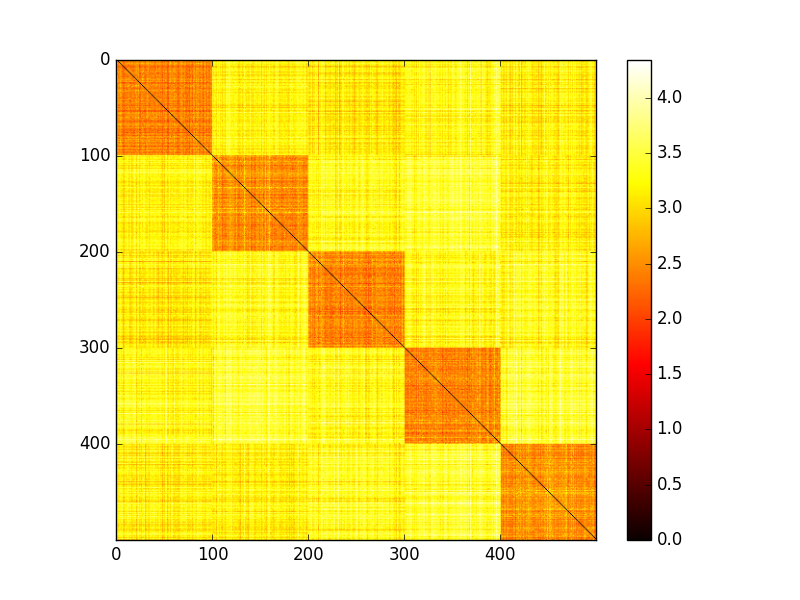}
  } &
  \subfloat[64 processors]{\label{fig:c4}%
  \includegraphics[width=4.3cm]{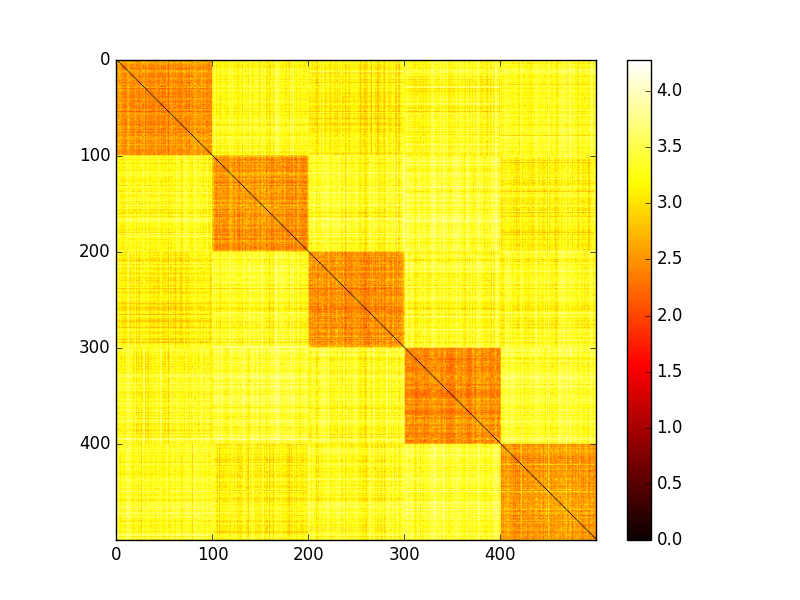}
  } \\
\end{tabular}
\caption{Distance matrix of the first 500 nodes of synthetic SBM networks based on the node embeddings produced by different number of processors. The color in the distance matrix indicates the distance between a pair of nodes, brighter the color longer the distance.}
\label{fig:tab1_fig}
\end{figure}

\subsection{Algorithm scalability}
We test our parallelization scheme on RPI Advanced Multiprocessing Optimized System (AMOS), which is a 5-rack, 5K nodes, 80K cores IBM Blue Gene/Q system~\cite{hari:ohm} with additional equipment. In AMOS supercomputer, each node consists of a 16-core, 1.6 GHz A2 processor, with 16 GB of DDR3 memory. Considering the fact that the inter-core communication is generally more efficient inside the same node than across different nodes, we use all the 16 cores of a node so that the communication between cores can be more efficient. In general, the maximum number of cores per node here is not constrained by the limit of 16 GB DDR3 memory space. This is one benefit of our memory management paradigm because every processor stores only one copy of the parameters of nodes and cascades associated with it.

\begin{figure}[t]
\centering
\begin{tabular}{c}
  \subfloat[m=100]{\label{fig:Speedup_SBM_v1_D100}%
  \includegraphics[width=7cm]{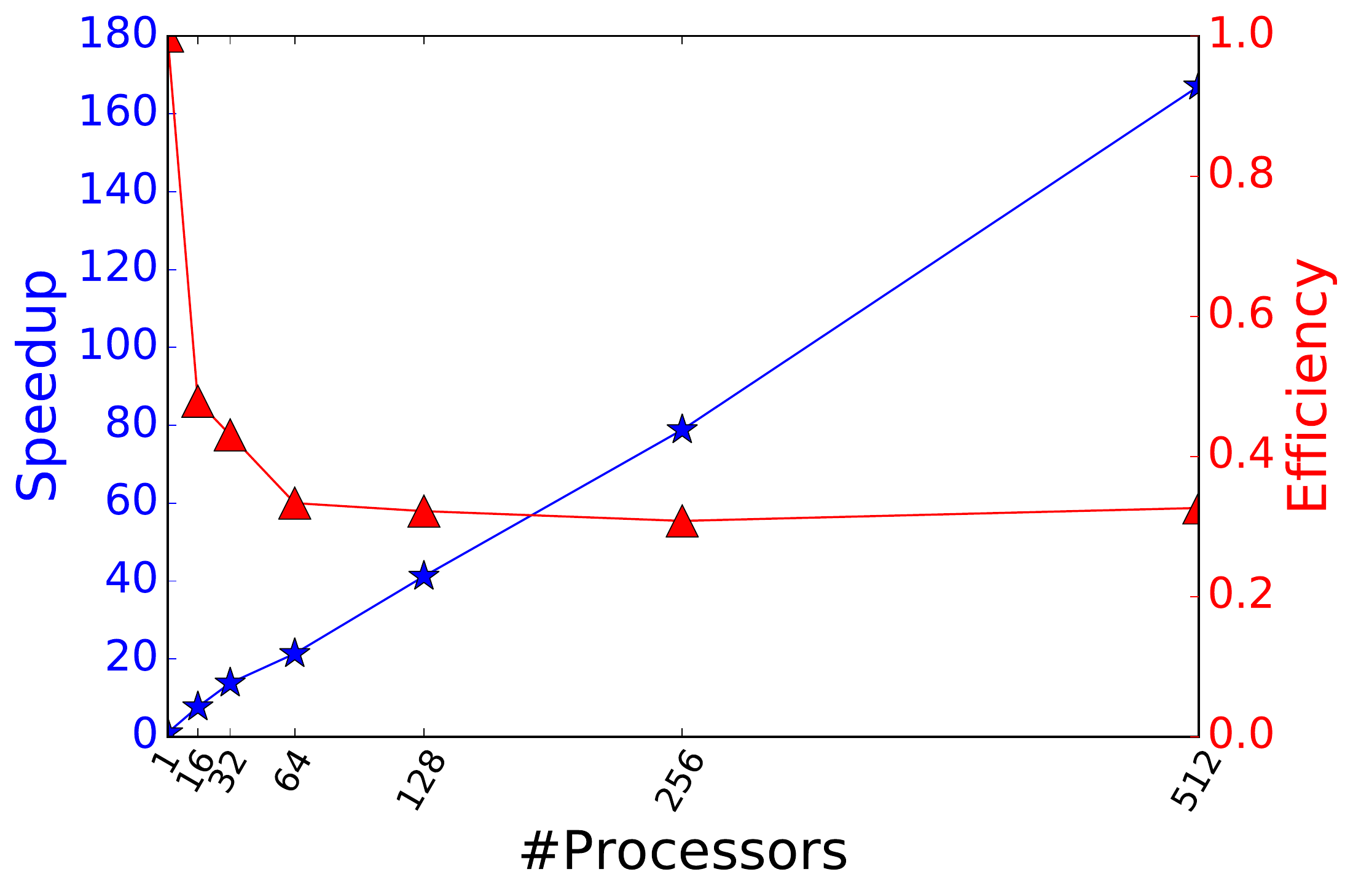}
  } \\
  \subfloat[m=200]{\label{fig:Speedup_SBM_v1_D200}%
  \includegraphics[width=7cm]{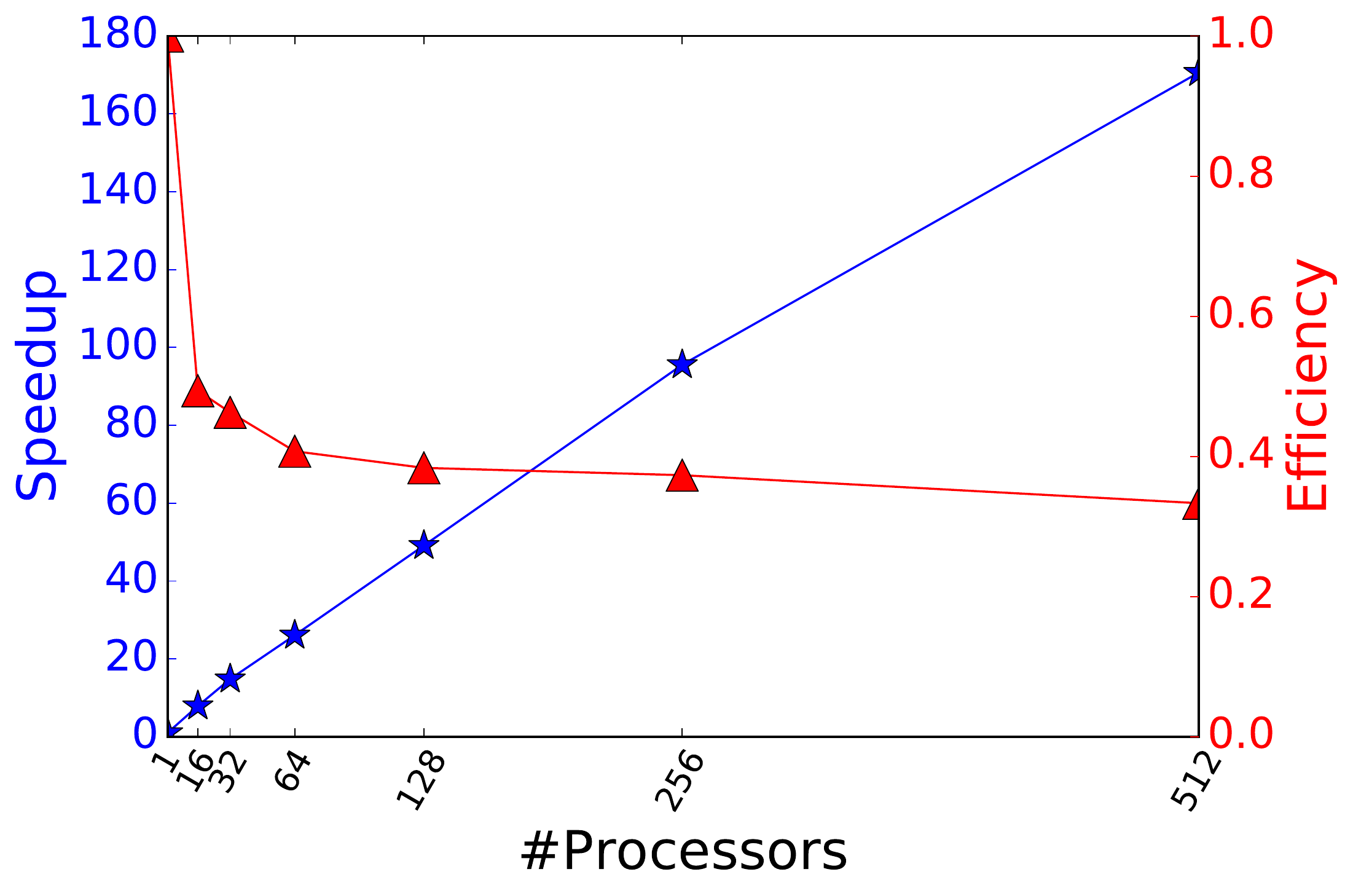}
  }
\end{tabular}
\caption{Speedup and efficiency of our parallelization scheme on Advanced Multiprocessing Optimized System (AMOS). $m$ denotes the dimension of $A_u$ and $M_c$.}
\label{fig:speedup_efficiency}
\end{figure}

%SBM_v1.txt 20K nodes, 10K cascades
The speedup and efficiency of our parallelization scheme are shown in Figure~\ref{fig:speedup_efficiency}. A total of 10K cascades are simulated on the SBM network with 20K nodes. Each cascade infects a total of 247 nodes on average. The Figure~\ref{fig:speedup_efficiency} shows that the parallelization scheme achieves an approximately linear speedup using several hundreds of processors. And the efficiency of the parallelization scheme is above 25\% in all cases. The comparison between Figure~\ref{fig:Speedup_SBM_v1_D100} and Figure~\ref{fig:Speedup_SBM_v1_D200} shows that the dimension $m$ does not change the speedup or efficiency. In our sampled GDELT dataset, the parallelized algorithm achieves the similar speedup and efficiency using 64 processors, but adding extra processors does not significantly increase the speedup due to the limited size of the dataset.

\begin{figure}[t]
\centering
\begin{tabular}{cc}
  \subfloat[]{\label{fig:node_number_time}%
  \includegraphics[width=3.72cm]{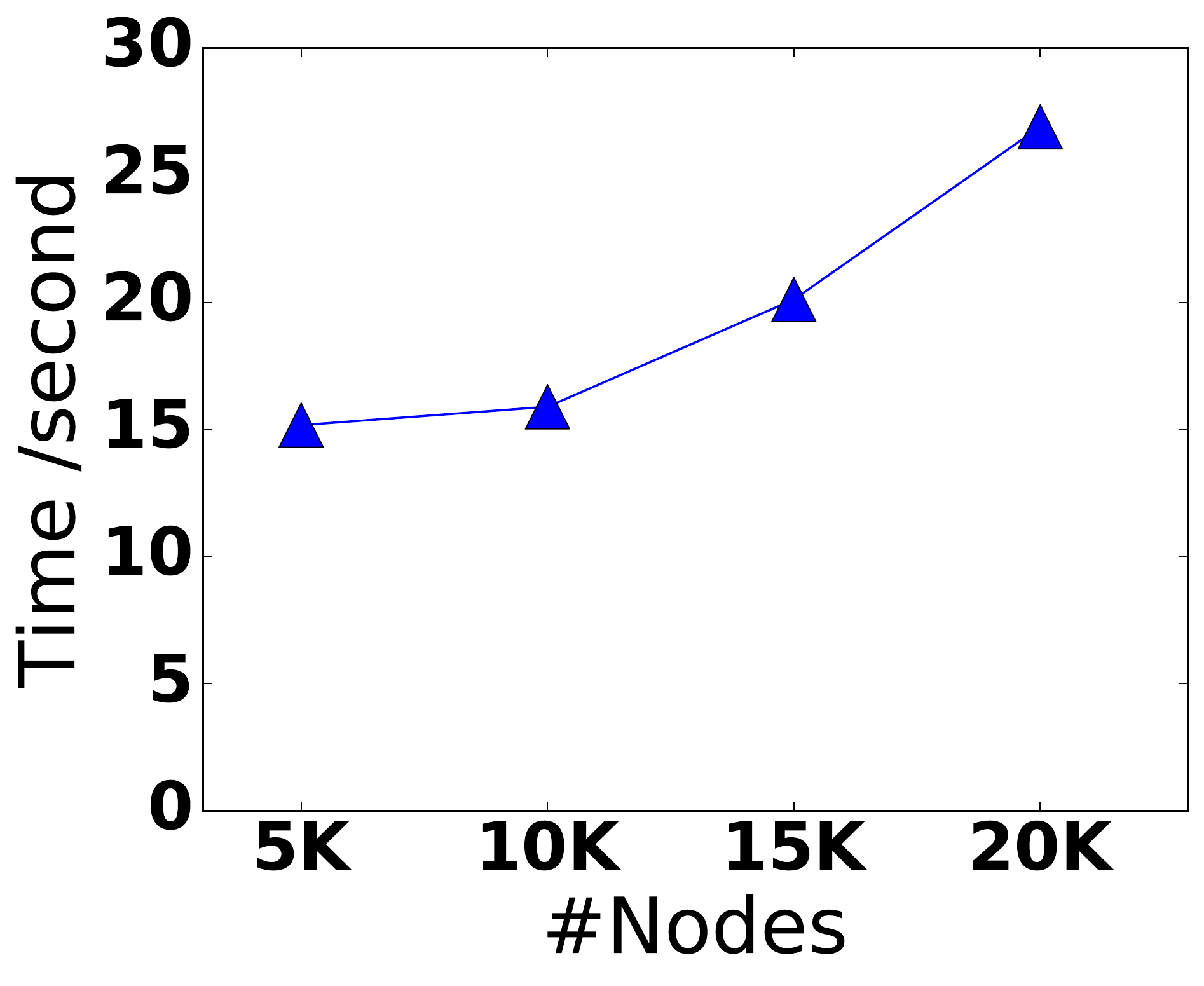}
  } &
  \subfloat[]{\label{fig:cas_number_time}%
  \includegraphics[width=3.72cm]{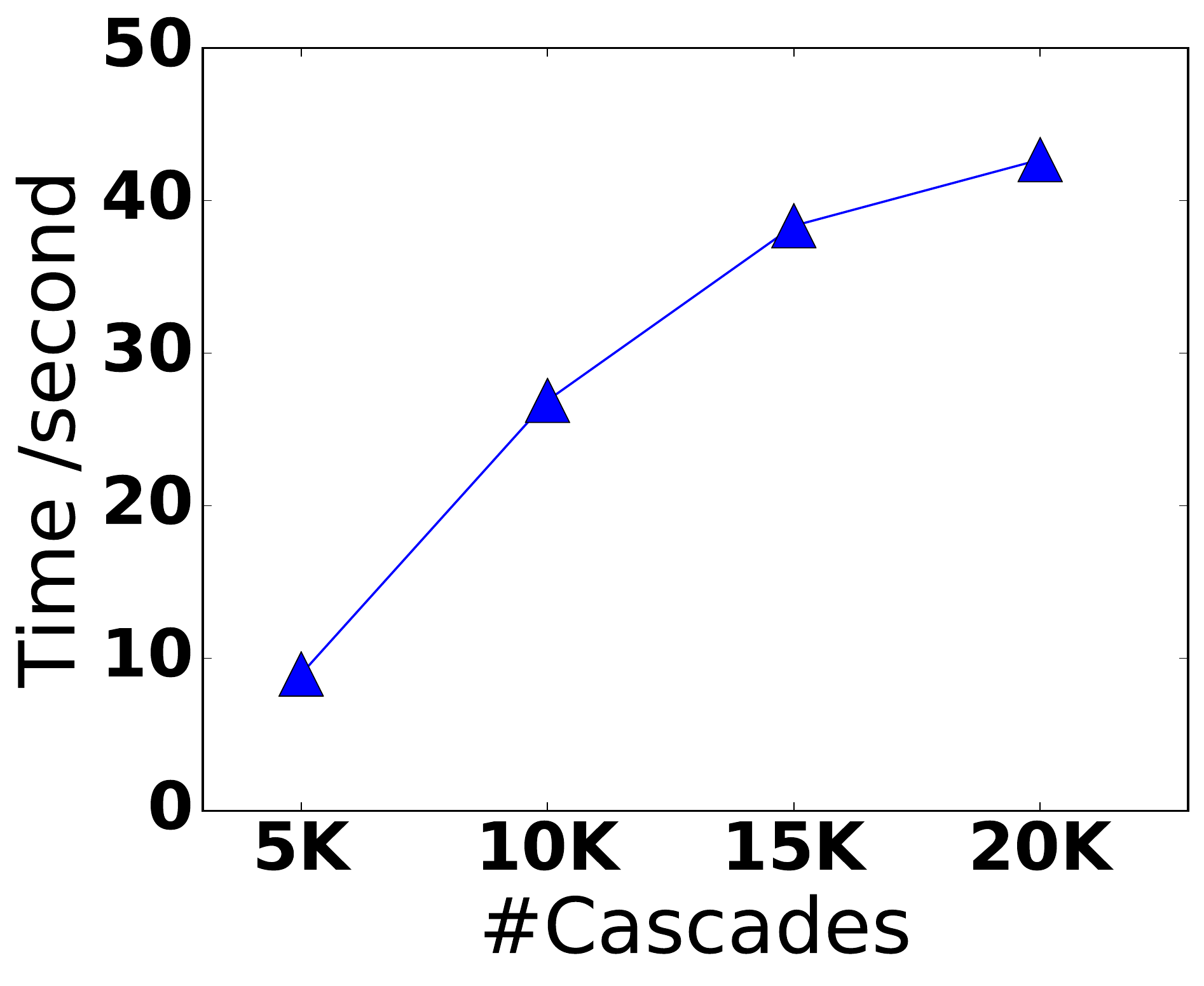}
  }\\
  \multicolumn{2}{c}{\subfloat[]{\label{fig:dimension_time}%
  \includegraphics[width=4.08cm]{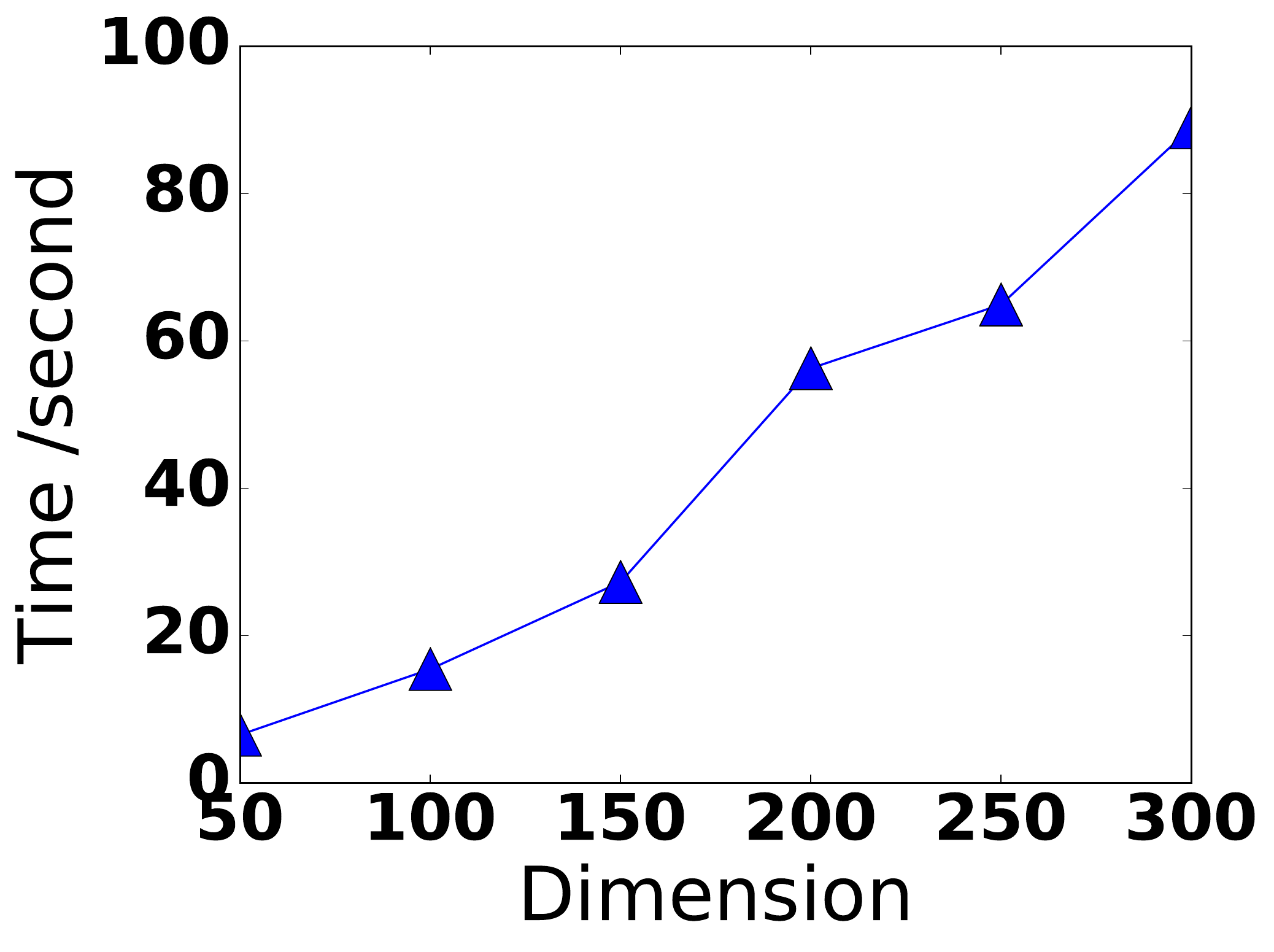}
  }}
\end{tabular}
\caption{The execution time of one SGA iteration in relation to the dimension $m$, the number of network nodes and the number of cascades.}
\end{figure}

To test the scalability of the parallelization scheme, we evaluate the execution time of one single SGA iteration of Algorithm~\ref{algo:2} using 512 processors. Given a network of 10K nodes and the different numbers of cascades, Figure~\ref{fig:node_number_time} shows that the execution time is approximately proportional to the number of cascades. A similar pattern is also observed as the dimension $m$ grows. In contrast, when the number of cascades is fixed and the number of network nodes increases, the execution time grows slowly - the execution time only doubles when the number of network nodes grows from 5K to 20K. The reason is that the number of infections per cascade is relatively stable in the synthetic data so that the increase of time for parameter propagation is actually smaller than the increase in the number of network nodes. In general, the parallelization scheme scales well with the number of network nodes, the number of cascades and the dimension $m$.

\begin{table*}[ht]
\caption{Accuracy of the news virality prediction measured by F1 scores. The predictions of our proposed model (ML) are consistently better than the the baseline model's (BL) using the same set of early adopters in the first 30 to 60 minutes.}
\centering
\label{tab:onehour}
\begin{tabular}{|c|c|c|c|c|c|c|c|c|c|c|c|c|}
\hline
\multirow{2}{*}{Threshold $\theta$} & \multicolumn{2}{c|}{$\tau=$0.5 Hour} & \multicolumn{2}{c|}{$\tau=$0.6 Hour} & \multicolumn{2}{c|}{$\tau=$0.7 Hour} & \multicolumn{2}{c|}{$\tau=$0.8 Hour} & \multicolumn{2}{c|}{$\tau=$0.9 Hour} & \multicolumn{2}{c|}{$\tau=$1 Hour} \\ \cline{2-13}
& BL & ML & BL & ML & BL & ML & BL & ML & BL & ML & BL & ML\\
\hline 
\hline
90\% &  0.410 & 0.500 & 0.410 & 0.497  & 0.410 & 0.499  & 0.499 & 0.549  & 0.499 & 0.548  & 0.534 & 0.594 \\  
\hline
91\% &  0.412 & 0.484 & 0.405 & 0.486  & 0.405 & 0.480  & 0.484 & 0.534  & 0.490 & 0.536  & 0.532 & 0.584 \\
\hline
92\% &  0.389 & 0.473 & 0.394 & 0.472  & 0.392 & 0.471  & 0.463 & 0.530  & 0.463 & 0.531  & 0.524 & 0.578 \\
\hline
93\% &  0.294 & 0.455 & 0.296 & 0.453  & 0.294 & 0.453  & 0.434 & 0.516  & 0.436 & 0.511  & 0.497 & 0.561 \\
\hline
94\% &  0.207 & 0.433 & 0.209 & 0.433  & 0.213 & 0.436  & 0.393 & 0.489  & 0.393 & 0.491  & 0.460 & 0.542 \\
\hline
95\% &  0.191 & 0.404 & 0.191 & 0.409  & 0.195 & 0.408  & 0.352 & 0.465  & 0.347 & 0.462  & 0.442 & 0.515 \\
\hline
96\% &  0.147 & 0.393 & 0.141 & 0.389  & 0.147 & 0.389  & 0.287 & 0.438  & 0.285 & 0.442  & 0.383 & 0.502 \\
\hline
97\% &  0.114 & 0.360 & 0.116 & 0.366  & 0.113 & 0.361  & 0.233 & 0.402  & 0.232 & 0.407  & 0.326 & 0.465 \\
\hline
98\% &  0.097 & 0.311 & 0.091 & 0.316  & 0.097 & 0.316  & 0.160 & 0.365  & 0.166 & 0.357  & 0.236 & 0.386 \\
\hline
99\% &  0.119 & 0.288 & 0.107 & 0.292  & 0.104 & 0.279  & 0.148 & 0.326  & 0.148 & 0.324  & 0.152 & 0.324 \\
\hline
\end{tabular}
\end{table*}

\subsection{Virality prediction} \label{set:virality}
Our aim is to predict the viral news cascades at their early stage. Specifically, with the global news dataset of events, the task is to predict the most reported events within a limited time period. Therefore, we rank the events reported in the news by the number of their reports and divide them into two classes: those who are among the top $(1-\theta)$ percent of this ranking and the remaining events. In this way, we can treat the virality prediction as a binary classification problem - given the early reports within a limited time period, can we classify the events reported in the news into these two categories? Since we are only interested in predicting the most viral events reported in the news, the threshold $\theta$ ranges from 90\% to 99\% in our experiments. Notice that a high threshold $\theta$ would result in two very imbalanced sets of samples, which would make the prediction challenging.

\textbf{Baseline} We build a baseline algorithm which uses multiple features extracted from cascade early progress and the Random Forest model~\cite{liaw:wie} for cascade classification. We choose Random Forest for comparison because, as an ensemble learning method, it is known to work well with the non-linear growth of modeled phenomena such as the viral spread of news reports. The extracted features include the number of unique early adopters, the frequency of the infections at the early stage, the maximum interval between two continuous infections, and the minimum interval between two continuous infections. In contrast, our proposed model uses the numbers of neighbors in different ranges from the infected nodes, i.e. vector $K$ presented in Eq.~\ref{eq:prediction_}, as the input of the Random Forest model. For a fair comparison, both the baseline and our model use the information about the early adopters in the first $\tau$ hours, where $\tau$ ranges from $0.5$ to $3$.

\begin{figure}
    \centering
    \hspace*{-1.5cm} 
    \includegraphics[width=9.5cm]{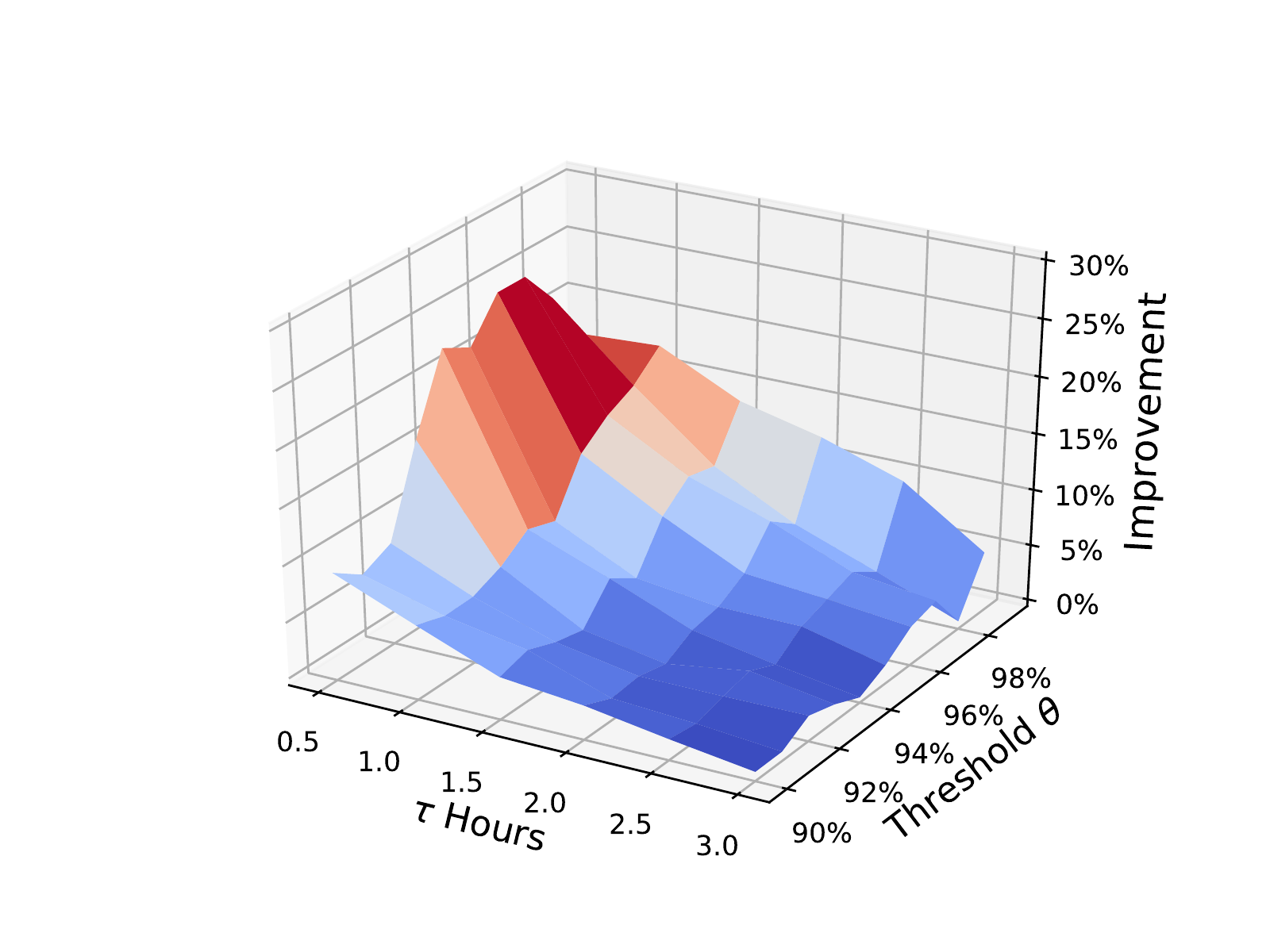}
    \caption{The improvement in virality prediction accuracy produced by our model in GDELT dataset in relation to the classification threshold $\theta$ and the initial observation period of $\tau$ hours.}
    \label{fig:improvement3d}
\end{figure}

The accuracy of the prediction is evaluated by the F1 score which is commonly used in the classification problems

\begin{figure*}[t]
\centering
\begin{tabular}{ccc}
  \subfloat[$\tau=0.5$ hours]{\label{fig:gdelt1}%
  \includegraphics[width=5.0cm]{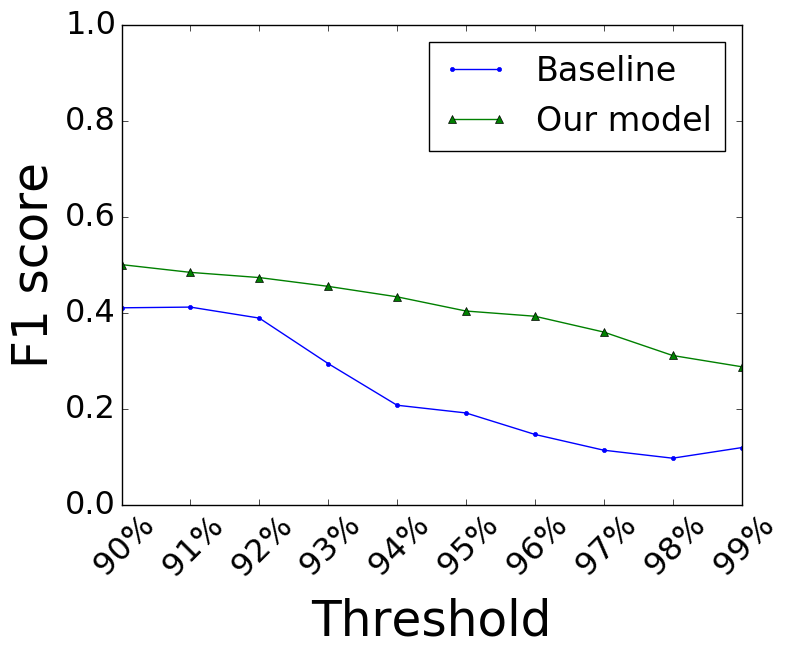}
  } &
  \subfloat[$\tau=1$ hours]{\label{fig:gdelt2}%
  \includegraphics[width=5.0cm]{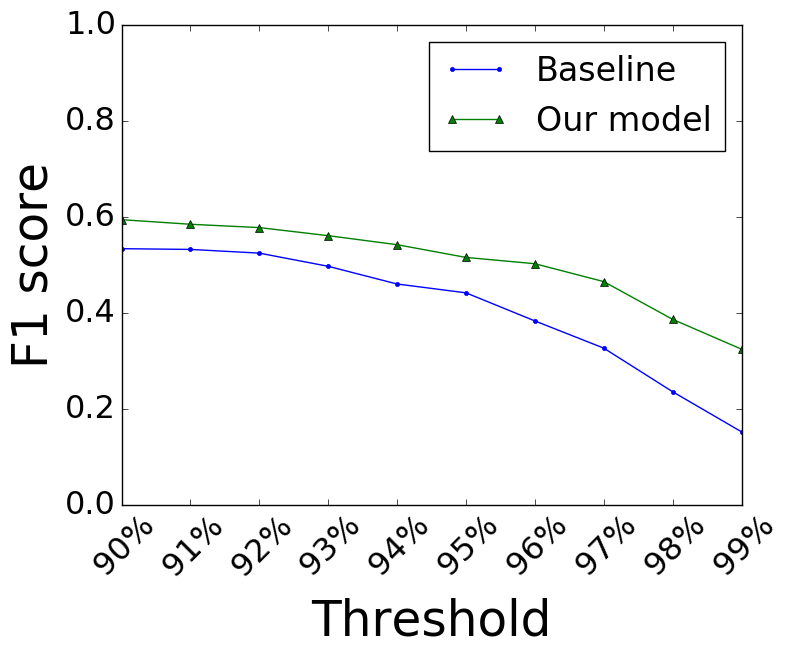}
  } &
  \subfloat[$\tau=1.5$ hours]{\label{fig:gdelt3}%
  \includegraphics[width=5.0cm]{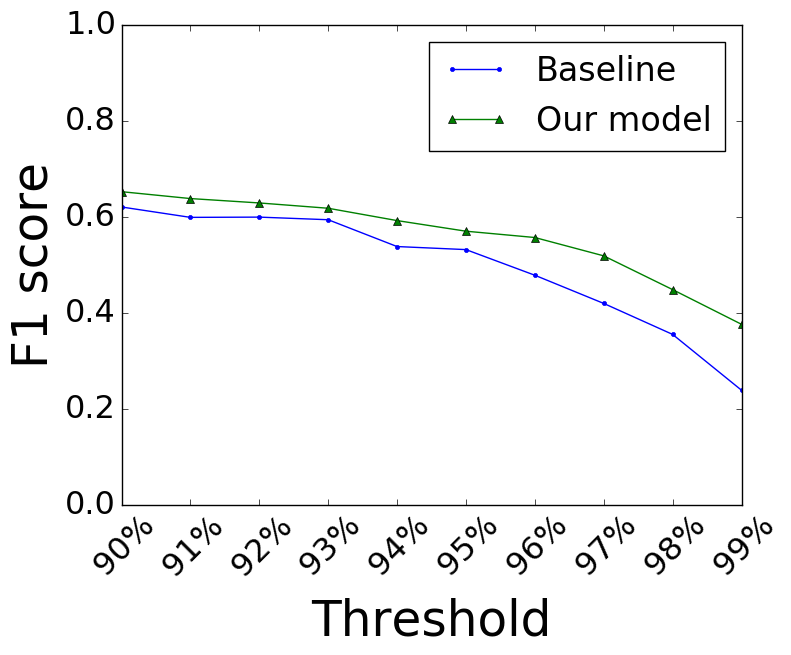}
  } \\
  \subfloat[$\tau=2$ hours]{\label{fig:gdelt4}%
  \includegraphics[width=5.0cm]{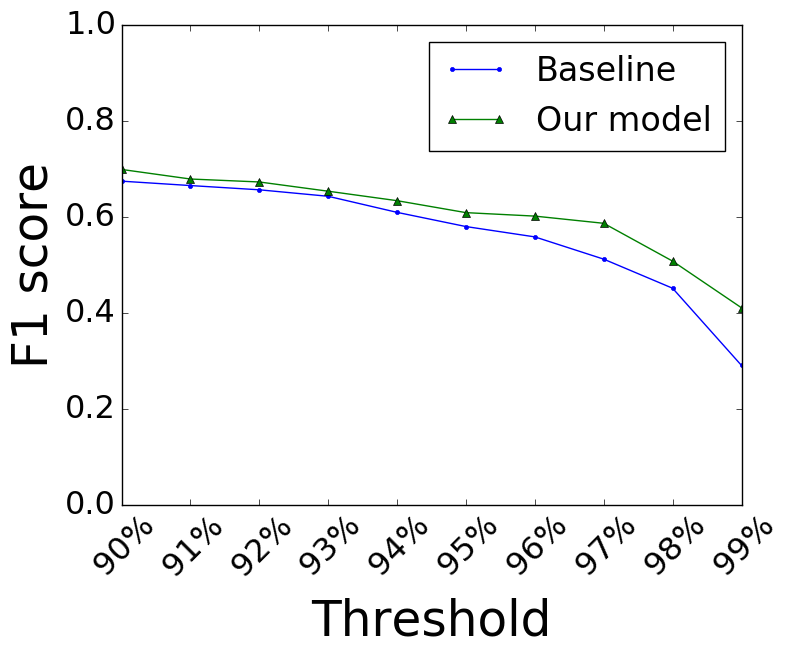}
  } &
  \subfloat[$\tau=2.5$ hours]{\label{fig:gdelt5}%
  \includegraphics[width=5.0cm]{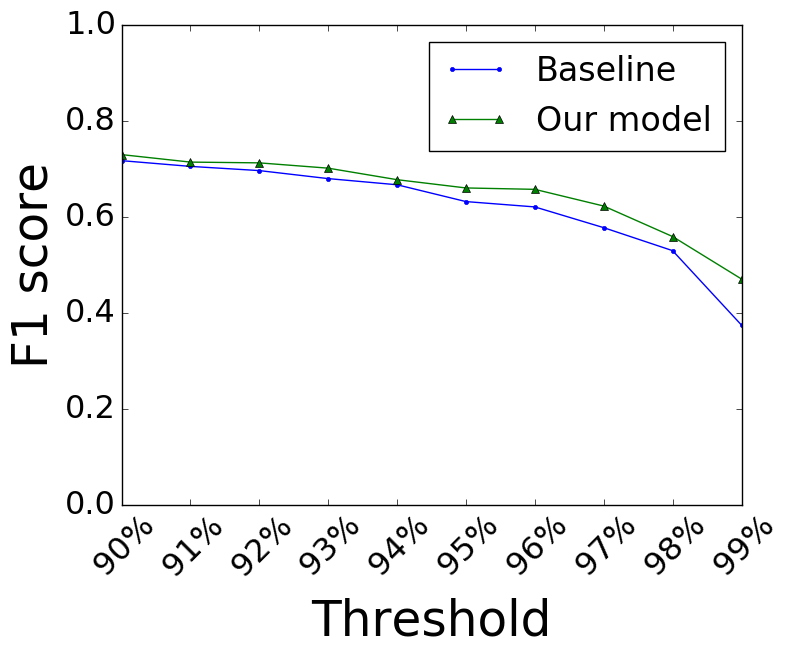}
  } &
  \subfloat[$\tau=3$ hours]{\label{fig:gdelt6}%
  \includegraphics[width=5.0cm]{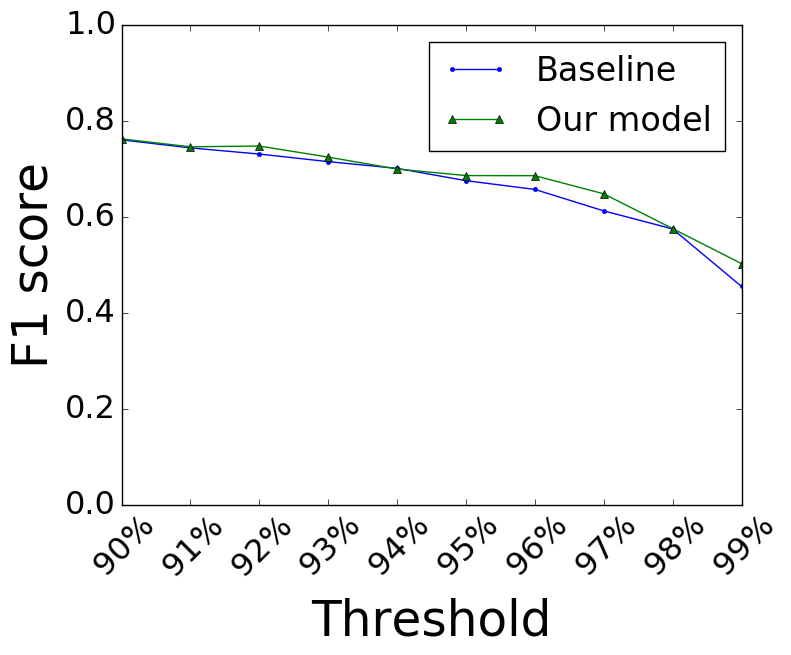}
  }
\end{tabular}
\caption{The accuracy of virality prediction in global news dataset of events (i.e. GDELT) measured by the F1 score. The prediction models take the news sites reporting an event in the first $\tau$ hours, i.e. the early adopters, as input.}
\label{fig:gdelt_prediction}
\end{figure*}

\begin{equation}
F_1 = \frac{2 \cdot \text{\it precision}\cdot \text{\it recall}}{\text{\it precision}+\text{\it recall}}
\end{equation}
F1 score considers both the precision and recall of virality prediction. A decent F1 score prevents the system from either predicting too many viral news with a high false positive rate, or from being too conservative making insufficient predictions.

Figure~\ref{fig:gdelt_prediction} shows the F1 scores of the 6-fold cross validation tests using a variety of $\tau$ values. As the threshold $\theta$ increases, the F1 scores of both the baseline and our model decrease due to the imbalanced sets of samples. The F1 scores of the prediction made by our model are consistently better than the baseline's. Specifically, our model outperforms the baseline by 10\% in most cases. As shown in Figure~\ref{fig:improvement3d}, the improvement produced by our model is very obvious when the value of $\tau$ becomes small. It is because our model uses the community structure of the propagation network which is not included in the feature-based baseline model. As the value of $\tau$ increases, both the baseline model and our proposed model gain better performance, yet the performance gap between them decreases at the same time. One potential reason is that the information cascades start to slow down inside each communities at this stage with $\tau>2$. Thus, the community structure does not provide extra signal for the prediction as it does at the early stage with $\tau<1.5$. In general, in terms of the prediction accuracy, the influence of $\tau$ value is more significant in the baseline model than in our model, which indicates that community structures can provide the critical signals to forecast the viral information cascades at the early stage.

The relationship between the improvement on prediction accuracy and the classification threshold $\theta$ is shown in Figure~\ref{fig:improvement3d}. Our proposed model performs much better than the baseline model when the threshold is high, and it achieves an almost $25\%$ improvement with the threshold $\theta=98\%$. As discussed in Section~\ref{sec:forecast}, our proposed method calculates the number of neighbors in the early adopters' the local neighborhood, i.e. the number of nodes whose $A_u$ vectors are close to the early adopters' in the latent space. Here, the most plausible explanation is that the most viral cascades have the early adopters in multiple dense areas so that they have advantages in disseminating the contagion to their neighbors in these regions in parallel, resulting in the viral infections within a limited time period. This explanation matches our observation about the viral news in online media - most news about events rarely cross the geographical and cultural boundaries, but once they do, the breaking news draw attention from the news media sites in different regions and hit the headlines very quickly.

\section{Conclusion}
We exploit the latent community structure in the global news network to improve the prediction of the viral cascades of news about events. The cascades which have early adopters in different communities have advantages in disseminating the contagion to these communities in parallel, and therefore are more likely to result in the viral infections within a limited time period. Our model captures such property by inferring the community structure using the response times of nodes rather than using the explicit network topology, because the references to propagation sources are usually missing in the real datasets. Due to the size of the relevant datasets, we successfully parallelized the inference algorithm for distributed memory machines and tested this parallelization on the RPI Advanced Multiprocessing Optimized System (AMOS) achieving orders of magnitude speedup.

% or


%\appendices

%\section{}

\ifCLASSOPTIONcompsoc
  \section*{Acknowledgments}
\else
  \section*{Acknowledgment}
\fi

\ifCLASSOPTIONcaptionsoff
  \newpage
\fi

This work was supported in part by the Army Research Laboratory (ARL) under Cooperative Agreement Number W911NF-09-2-0053, (NS CTA), by the Office of Naval Research (ONR) grant No. N00014-15-1-2640, and by the Army Research Office (ARO), grant W911NF-16-1-0524. The views and conclusions contained in this document are those of the authors and should not be interpreted as representing the official policies either expressed or implied of the Army Research Laboratory, or the U.S. Government.



%\bibliographystyle{IEEEtran}

\vfill


\end{document}